\documentclass[reqno]{amsart}

\usepackage{mathrsfs}
\usepackage{dsfont}
\usepackage[hypertex]{hyperref}  

\newcommand{\R}[0]{\mathds{R}}
\newcommand{\C}[0]{\mathds{C}}
\newcommand{\N}[0]{\mathds{N}}

\newtheorem{theorem}{Theorem}[section]
\newtheorem{corollary}[theorem]{Corollary}
\newtheorem{lemma}[theorem]{Lemma}
\newtheorem{proposition}[theorem]{Proposition}
\newtheorem{as}{Assumption}
\newtheorem{con}{Condition}
\newtheorem{remark}[theorem]{Remark}
\newtheorem{definition}[theorem]{Definition}

\newcommand{\eref}[1]{(\ref{#1})}

\newcommand{\ol}[1]{\overline{#1}}
\newcommand{\ul}[1]{\underline{#1}}
\newcommand{\ep}[0]{\varepsilon}

\numberwithin{equation}{section}

\begin{document}

\title[complex geometric optics for symmetric hyperbolic systems I]{complex geometric optics for symmetric hyperbolic systems I: linear theory}

\author{Omar Maj}

\address{Max-Planck-Institut f\"ur Plasmaphysik, D-85748 Garching, Germany}

\email{omaj@ipp.mpg.de}

\begin{abstract}
We obtain an asymptotic solution for $\ep \to 0$ of the Cauchy problem for linear first-order symmetric hyperbolic systems with oscillatory initial values written in the eikonal form of geometric optics with frequency $1/\ep$, but with complex phases. For the most common linear wave propagation models, this kind on Cauchy problems are well-known in the applied literature and their asymptotic theory, referred to as complex geometric optics, is attracting interest for applications. In this work, which is the first of a series of papers dedicated to complex geometric optics for nonlinear symmetric hyperbolic systems, we develop a rigorous linear theory and set the basis for the subsequent nonlinear analysis.
\end{abstract}


\maketitle

\section{Introduction}
\label{intro}

We shall obtain an asymptotic solution for $\ep \to 0$ of the Cauchy problem for linear first-order symmetric hyperbolic systems in several spatial dimensions with initial values in the form, 
\begin{equation*}
h^\ep(x) = \sum_{\mu =1}^m h_\mu (x) e^{i\psi_\mu(x)/\ep}, \quad \big(\text{$\psi_\mu(x)$ are complex-valued phases}\big),
\end{equation*}
that is, a superposition of waves of frequency $1/\ep$, amplitude $h_\mu$ and complex phase $\psi_\mu$ for which we always assume the condition $\mathrm{Im} (\psi_\mu)\geq 0$. 

More specifically, we construct a family of waves $v^\ep$ such that $|h^\ep -v^\ep_{|t=0}| = O(\ep^{\frac{1}{2}})$ and $|L(t,x,\partial) v^\ep | = O(\ep^{\frac{1}{2}})$ where $L(t,x,\partial)$ is the considered symmetric hyperbolic system; then, we prove the convergence of the asymptotic solution $v^\ep$ to the exact one in a suitable topology. 

The strategy will be introducing complex-valued phase functions in the framework of geometric optics for which we refer to the lectures by Rauch \cite{R} where rigorous results are given together with a comprehensive list of references. In the applied literature [2-6] the theory of oscillatory waves with a complex-valued phase is widely developed in several different variants, cf., the recent book by Kravtsov \cite{Kr} for a tentative classification; particularly, the beam tracing method of Pereverzev \cite{P96,P98} is currently used in fusion experiments and complex geometric optics finds several applications in electrodynamics and geophysics \cite{Kr, KrBe}. On the other hand, to our knowledge, a rigorous justification of such approaches is still lacking even in the linear case, especially for systems, and no attempts have been made to address the nonlinear case. Here, we put the complex geometric optics theory on a rigorous basis by means of novel tools that clarify the analysis of the polarization in the presence of complex phases as well as by a precise argument based on a partition of unity. Although it may be of independent interest, this result is meant to be the first part of a work dedicated to the study of such initial value problems for nonlinear symmetric hyperbolic systems.

The main point is that the non-negative imaginary part of the phase yields a strong localization of $h^\ep$ around the set $R^o = \bigcup_\mu \{x;\ \mathrm{Im}\psi_\mu(x)=0\}$, as $\ep \to 0$, thus, $h^\ep$ exhibits features common to both wave trains (oscillatory character) and wave pulses (localization): the more the wave field is localized, the more it oscillates in such a way that, qualitatively speaking, the number of oscillations under the wave field envelope can be thought of to be independent on $\ep$. For such localized waves, \emph{diffraction effects} should be taken into account in the leading order asymptotics, even for short-time propagation. Our analysis parallels the results on short pulses \cite{AR1,AR2}, albeit diffraction is accounted for in a different way, i.e., through the coupling of the real and imaginary parts of the phases. Such a description of diffraction should be compared to a deformation of the wave field envelope described by a Schr\"odinger-type equation relevant to the paraxial approximation (also known as ``parabolic wave equation'' \cite{BB}). In the forthcoming papers we shall see that this duality between complex phases and ``Schr\"odinger-type'' evolution of the wave field envelope is still valid in the nonlinear context. 

\section{General Assumptions and Main Results}
\label{results}

We shall consider systems of first-order partial differential equations of the form
\begin{equation}
\label{L}
L(t,x,\partial)u(t,x) = \partial_t u(t,x) + \sum_{j=1}^d A_j(t,x)\partial_{x_j} u(t,x) + B(t,x)u(t,x)=0,
\end{equation}
for the wave field $u \in C^\infty(\ol{\Omega}; \C^N)$ on the closure $\ol{\Omega}$ of an open, connected and bounded set $\Omega \subseteq \R^{1+d}$; these are the systems considered by Lax in his seminal paper \cite{Lax}. Here, $A_j$, $B$ are functions valued in the space $\mathrm{End}(\C^N)$ of $N\times N$ complex matrices depending smoothly on $(t,x)$.

\begin{as}
\label{hyperbolicity}
The matrices $A_j(t,x)$ are Hermitian in a neighbourhood of $\ol{\Omega}$. For the particular case of a single equation ($N=1$), this reduces to $A_j(t,x) \in \R$. 
\end{as}

Assumption \ref{hyperbolicity} is equivalent to requiring that the matrix 
\begin{equation*}
A(t,x,\xi) = \sum_{j=1}^d  A_j(t,x)\xi_j,
\end{equation*}
is Hermitian for $(t,x)$ in a neighbourhood of $\ol{\Omega}$ and $\xi \in \dot{\R}^d$, with $\dot{\R}^d = \R^d\setminus \{0\}$. This implies that the principal part of the operator \eref{L}, namely,
\begin{equation}
\label{L0}
L_0(t,x,\partial) = \partial_t + \sum_{j=1}^d A_j(t,x)\partial_{x_j},
\end{equation}
is strictly hyperbolic in time, that is, the characteristic equation $\det \sigma_{L_0}=0$, with
\begin{equation*}
\sigma_{L_0}(t,x,\tau,\xi) = i\big(\tau I + A(t,x,\xi)\big),
\end{equation*}
being the principal symbol of $L_0$, has real-valued roots given by the solution of
\begin{equation*}
f_l(t,x,\tau,\xi)=\tau + \lambda_l(t,x,\xi)=0,
\end{equation*}
with $\lambda_l$ the $l$-th eigenvalue of $A$. This defines a conic variety $\mathrm{Char}(L)$ in $T^*\R^{1+d}\setminus 0$, the cotangent bundle without the zero section, which is called characteristic variety or local dispersion surface in the physics literature; we will assume that different branches of the characteristic variety corresponding to different eigenvalues $\lambda_l$ cannot merge; specifically, we have the following assumption.

\begin{as}
\label{modes}
The eigenvalues $\lambda_l(t,x,\xi)$ have constant multiplicity.
\end{as}

We know that eigenvalues of constant multiplicity are everywhere distinct, thus, one can label them so that $\lambda_1 < \lambda_2 < \cdots < \lambda_l < \cdots$, and, through the implicit function theorem, one can show that they depend smoothly on the entries of the matrix $A$ \cite{AMR}; in addition, they are homogeneous of degree one, as follows by inspection of the characteristic equation. Moreover, there is a constant $r>0$ such that $|\lambda_l(t,x,\hat{\xi}) - \lambda_{l'}(t,x,\hat{\xi}) | \geq 2r$ for $(t,x,\hat{\xi}) \in \ol{\Omega} \times S^{d-1}$, hence, we can use the contour integral representation for the spectral projectors, \cite{R, Ka},
\begin{equation}
\label{spectral}
\pi_l(t,x,\hat{\xi}) = \frac{1}{2\pi i}\oint_{|z-\lambda_l(t,x,\hat{\xi})|=r} \big(zI - A(t,x,\hat{\xi})\big)^{-1}dz,
\end{equation}
which shows that $\pi_l(t,x,\xi)= \pi_l(t,x,\hat{\xi})$ are of class $C^\infty (\ol{\Omega}\times \dot{\R}^d)$ and homogeneous of degree $0$ in the variable  $\xi=|\xi|\hat{\xi}$. Finally we give the definition of complex phase.

\begin{definition}[Complex phase]
\label{complex_phase}
For $m \geq 0$ integer, $\phi \in C^\infty(\ol{\mathcal{O}}; \C^m)$ in a bounded open set $\mathcal{O} \subset \R^n$ is a complex phase iff, for each component $\phi_\mu = \varphi_\mu +  \chi_\mu$, one has $d\varphi_\mu \not =0$ and $\chi_\mu \geq 0$ in $\ol{\mathcal{O}}$.
\end{definition}

Now we choose the domain $\Omega$ according to  
\begin{equation*}
\ol{\Omega} = \{(t,x)\in\R^{1+d}; 0 \leq t \leq T, |x - \underline{x}| \leq \rho - ct \}, 
\end{equation*}
for suitable constants $\underline{x}\in \R^d$, $\rho > 0$, $T \in (0,\rho /c)$, with $c>0$ being the propagation speed of the operator \eref{L}; we recall that such an $\Omega$ is a domain of determinacy for the operator $L$ and the propagation speed $c$ is characterized by the condition, \cite{R},
\begin{equation*}
cI + \sum_{j=1}^d \frac{x_j}{|x|} A_j(t,x) \geq 0,
\end{equation*}
on the boundary of $\ol{\Omega}$; we denote $\ol{X^{t'}} = \ol{\Omega} \cap \{t=t'\}$, for $t' \in [0,T]$, with $X^t$ the interior of $\ol{X^t}$. Then, we consider the Cauchy problem for equation \eref{L} with initial value
\begin{equation}
\label{Cauchy-linear}
u^\ep_{|t=0}(x) = h^\ep(x) = \sum_{\mu=1}^m h_\mu(x) e^{i\psi_\mu(x)/\ep}, \qquad x \in \ol{X^o},
\end{equation}
where $\ep \in \R_+$, the functions $\psi = (\psi_\mu) \in C^\infty(\ol{X^o};\C^m)$ are complex phases and $h_\mu \in C^\infty(\ol{X^o};\C^N)$ are called amplitudes. We require that the initial complex phases fulfill the following geometric hypothesis. 

\begin{as}
\label{Rzero}
The locus $\mathrm{Im} \psi_\mu(x)=0$ amounts to a closed submanifold $R^o_\mu \subseteq X^o$ without boundary and $R^o_\mu \cap R^o_\nu = \emptyset$ for $\mu \not = \nu$ so that $R^o = \bigcup_\mu R^o_\mu$ is also a closed submanifold without boundary. 
\end{as}

As for the amplitudes $h_\mu \in C^\infty(\ol{X^o};\C^N)$, we assume the following condition which can always be satisfied after splitting each $h_\mu$ as appropriate.  

\begin{con}
\label{pol}
For every $\mu$ there is $l=l(\mu)$ such that
\begin{equation*}
\pi_l(0,x,d\mathrm{Re} \psi_\mu(x))h_\mu(x) = h_\mu(x), 
\end{equation*}
which means that each component of the initial datum is polarized in the eigenspace of a specific eigenvalue $\lambda_l$. Such specially polarized waves are referred to as eigenmodes in the applied literature.
\end{con}

We search for an asymptotic solution $v^\ep \in C^\infty(\ol{\Omega};\C^N)$ in the form
\begin{equation}
\label{cgo}
v^\ep(t,x) = \sum a^\ep_\mu(t,x) e^{i\phi_\mu(t,x)/\ep}, \quad a^\ep_\mu(t,x) = a_{\mu}^{(0)}(t,x) + \ep a_{\mu}^{(1)}(t,x),
\end{equation}
with $\phi = (\phi_\mu)_\mu$ being a multi-valued complex phase and $a_{\mu}^{(j)}$, $j=0,1$, are the amplitudes. Then, we readily get
\begin{multline}
\label{eq}
L(t,x,\partial) v^\ep = \sum \Big[\ep^{-1} \sigma_{L_0}(t,x,d\phi_\mu) a^{(0)}_\mu + \sigma_{L_0}(t,x,d\phi_\mu) a^{(1)}_\mu \\
+ L_0(t,x,\partial) a^{(0)}_\mu + B(t,x) a^{(0)}_\mu + \ep L_0(t,x,\partial)a^{(1)}_\mu +\ep B(t,x)a^{(1)}_\mu \Big]e^{i\phi_\mu/\ep},
\end{multline}
where the principal symbol $\sigma_{L_0}$ is a polynomial in $(\tau,\xi)$ and it extends to an entire function in such variables, hence, $\sigma_{L_0}(t,x,d\phi_\mu)$ makes sense for $\phi_\mu$ complex-valued.

Usually, one imposes in \eref{eq} that the coefficients of different powers of $\ep$ vanish identically. Here, instead, the presence of the exponential $e^{- \chi_\mu/\ep}$, $\chi_\mu=\mathrm{Im} \phi_\mu$, allows us to consider weaker conditions; precisely we have the following estimates used independently by Maslov \cite{Ma} and Pereverzev \cite{P96,P98}.

\begin{lemma}
\label{maslov-estimate1}
Let $k\geq 0$ be an integer, $\phi =\varphi + i \chi \in C^\infty (\ol{\Omega};\C)$, $f \in C^\infty(\ol{\Omega})$ and let $S \subset \ol{\Omega}$ be any (non-empty) set such that $\ol{S} \cap \{\chi(t,x)=0\} = \emptyset$. Then,
\begin{equation*}
\ep^{-k}|f e^{i\phi/\ep}| \leq C_{k}, \quad \text{for every $(t,x) \in \ol{S}$ and $\ep \in\R_+$},
\end{equation*}
where $C_k = k^k e^{-k} \sup_{(t,x)\in \ol{S}} |f(t,x) /\chi^{k}(t,x)|$.
\end{lemma}

\begin{proof}
Since $\chi >0$ in $\ol{S}$, the function $\ep \mapsto \ep^{-k} e^{-\chi/\ep}$ is bounded for every $(t,x)\in \ol{S}$ with maximum value $(k/e\chi)^k$.
\end{proof}

Hence, $v^\ep$ and $Lv^\ep$ are localized around the set
\begin{equation*}
R = \bigcup_{\mu=1}^m R_\mu, \quad \text{with}\quad R_\mu = \{(t,x)\in \ol{\Omega}; \chi_\mu (t,x) =\mathrm{Im}\phi_\mu(t,x)=0\};
\end{equation*}
both $R$ and $R_\mu$ are assumed to be submanifold of $\ol{\Omega}$ characteristic for the operator \eref{L} and referred to as the \emph{reference manifold} for the complex phases $\phi$ and $\phi_\mu$, respectively.

\begin{as}
\label{ref-man}
In a conic neighbourhood of $\ol{\Omega}$ in $T^*\R^{1+d}\setminus 0$, there are isotropic submanifolds $\Lambda_1, \ldots, \Lambda_m$ such that, for $\mu \in \{1,\ldots,m\}$,
\begin{itemize}
\item[(i)] $(0,x,\tau,\xi)\in \Lambda_\mu$ only if
\begin{equation*}
x\in R^o_\mu, \quad  \xi = d\psi_\mu(x),
\end{equation*}
with $R^o_\mu$ given in assumption \ref{Rzero} and $\psi_\mu$ given in the initial value \eref{Cauchy-linear}; when $x\in R^o_\mu$, $d\psi_\mu(x)$ is real valued;
\item[(ii)] with $l(\mu)$ given in condition \ref{pol}, and for every $(t,x,\tau,\xi)\in \Lambda_\mu$,
\begin{equation*}
f_{l(\mu)}(t,x,\tau,\xi)=0, \ \text{and}\  H_{f_{l(\mu)}} \in T_{(t,x,\tau,\xi)}\Lambda_\mu,
\end{equation*}
where $H_f$ is the Hamiltonian field corresponding to the Hamiltonian function $f$;
\item[(iii)] the canonical projection $T^*\R^{1+d} \to \R^{1+d}$ restricts to diffeomorphisms between $\Lambda_\mu \cap \{(t,x,\tau,\xi); (t,x)\in \ol{\Omega}\}$ and a closed submanifold $R_\mu \subset \ol{\Omega}$; $\mathrm{dim} R_\mu = \mathrm{dim}R_\mu^o +1$ and $R_\mu$ is transverse to $X^T$; 
\item[(iv)] the submanifolds $R_1,\ldots, R_m$ are disjoint.
\end{itemize}
\end{as}

\begin{remark}
\label{caustics}
The construction of the isotropic submanifolds $\Lambda_\mu$ is exactly the same as in the standard geometric optics method, the only difference being its dimension. With minor modification we can also consider the (degenerate) case in which $R^o_\mu$ amounts to a single point and, correspondingly, $R_\mu$ is a single geometric optics ray; this is the case considered in most applications [3-6].
\end{remark}

Assumption \ref{ref-man} addresses the two main geometric objects of the theory, that is, the isotropic manifold $\Lambda_\mu$ and the corresponding projection $R_\mu$. In the theory of Maslov \cite{Ma} the focus is on the former, whereas in the paraxial approach by Pereverzev \cite{P96,P98} one looks at the latter. In section \ref{gcee}, we shall see that assumption \ref{ref-man} is, indeed, an hypothesis on the characteristics flow for the operator \ref{L}. It allows us, in particular, to construct a coordinate chart $(\mathcal{O}_\mu, \kappa_\mu)$, where $\mathcal{O}_\mu$ is a neighbourhood in $\ol{\Omega}$ of a point $(t_0,x_0) \in R_\mu$ and $\kappa_\mu : \mathcal{O}_\mu \ni (t,x) \mapsto (t,r,s) \in  [0,T] \times \mathcal{O}_r \times \mathcal{O}_s$ is a diffeomorphism, with $\mathcal{O}_r \subseteq \R^{d_1}$, $\mathcal{O}_s \subseteq \R^{d_2}$ ($d_1+d_2=d$), so that $\mathcal{O}_\mu \cap R_\mu$ is mapped into $\{s=0\}$; the latter is just the submanifold property \cite{AMR} and we see that the boundary $\partial R_\mu$, corresponding to $t=0$ and $t=T$, is plainly accounted for.  

In addition, in section \ref{gcee}, we shall see that $\chi_\mu(t,x) = \mathrm{Im}\phi_\mu(t,x) \geq c |s|^q$ for a constant $c>0$ and an even integer $q>0$.

\begin{lemma}
\label{maslov-estimate2-small}
Let $\phi$ and $f$ be as in lemma \ref{maslov-estimate1} and let $R = \{\mathrm{Im}\phi = \chi=0\}$ be a submanifold admitting coordinates $(t,r,s) = \kappa(t,x)$ on a neighbourhood $\mathcal{O}\subset \ol{\Omega}$ as described above. We assume further that $\chi(t,r,s) \geq c |s|^q$ in $[0,T]\times \mathcal{O}_r \times \mathcal{O}_s$ where $c>0$ and $q \in \dot{\N}$ is an even integer. Then, for every $k \in \N$, $t \in [0,T]$ and $A \subset \mathcal{O}_r \times \mathcal{O}_s$ open and bounded with $A \cap \{s=0\}\not = \emptyset$, there are constants $C_{\alpha}$, $\alpha\in \N^{d_2}$ with length $|\alpha|=k$, such that
\begin{equation*}
\big| \big(f(t,r,s) - \sum_{|\alpha| < k} c_\alpha(t,r) s^\alpha\big)e^{i\phi(t,r,s)/\ep}\big| \leq \ep^{\frac{k}{q}} \sum_{|\alpha|=k}\ \sup_{(r,s)\in \ol{A}} |\partial^\alpha_s f(t,r,s)|  C_{\alpha},
\end{equation*}
uniformly for $(r,s) \in \ol{A}$, $\ep \in \R_+$; such an estimate can be made uniform in $[0,T]\times \ol{A}$. Here, $f(t,r,s) = f\circ\kappa^{-1}(t,r,s)$ and analogously for the other functions, whereas $c_\alpha(t,r) = \partial^\alpha_s f(t,r,0)/\alpha!$.
\end{lemma}

\begin{proof}
First, let us fix a time $t\in[0,T]$. By Taylor formula we have
\begin{equation*}
f(t,r,s) - \sum_{|\alpha| < k} c_\alpha(t,r) s^\alpha = \sum_{|\alpha| = k} s^\alpha \tilde{c}_\alpha(t,r,s),
\end{equation*}
where $|\tilde{c}_\alpha(t,r,s)| \leq \sup_{\ol{A}} |\partial_s^\alpha f(t,r,s)|$ in $\ol{A}$. Moreover, if $|\alpha|=k$,
\begin{equation*}
|s^\alpha e^{i\phi/\ep}| \leq |s^\alpha e^{-c |s|^q /\ep}| = \ep^{\frac{k}{q}} |v^{\alpha} e^{-c |v|^q}|\leq \ep^{\frac{k}{q}} C_{\alpha},
\end{equation*}
with $C_{\alpha}$ being the maximum of $|v^\alpha| e^{-c|v|^q}$ for $v \in \R^{d_2}$. This yields the claimed estimate pointwise in $[0,T]$ which, on the other hand, entails the uniform estimate in $[0,T]\times\ol{A}$ as the right-hand side is continuous on $[0,T]$.
\end{proof}

In virtue of lemma \ref{maslov-estimate1} we see that it is enough to consider a neighbourhood of the reference manifold $R$ and lemma \ref{maslov-estimate2-small}, specialized for $\phi = \phi_\mu$, tells us that it is enough to find a solution of
\begin{align}
\label{eq1}
&\sigma_{L_0}(t,x,d\phi_\mu) a^{(0)}_\mu = O(|s|^{\frac{3q}{2}}),\\
\label{eq2}
&\sigma_{L_0}(t,x,d\phi_\mu) a^{(1)}_\mu + L_0(t,x,\partial)a^{(0)}_\mu + B(t,x)a^{(0)}_\mu= O(|s|^{\frac{q}{2}}),
\end{align}
in a neighbourhood of $R_\mu$. The lower non-trivial value of the parameter $q$ is clearly $q=2$ and this is the case we are interested in. 

The first result of the paper is the existence of an equivalence class of solutions to equations \eref{eq1} and \eref{eq2}. This can be done by means of standard tools: an extension argument put forward in section \ref{sigma}, the construction of an approximate solution for the so-called complex eikonal equation for addressed in section \ref{gcee} together with the analysis of equation \eref{eq1} and \eref{eq2} in section \ref{amplitudes-linear}. The construction of the complex geometric optics solution is completed in section \ref{proof1} where the following proposition and its corollary are proved.

\begin{proposition}
\label{main-linear}
Let assumptions \ref{hyperbolicity}-\ref{ref-man} be satisfied together with condition \ref{pol} and $d_s^2 \mathrm{Im}\psi_\mu(x) > 0$ when $x \in R_\mu^o$. Then, there exists an equivalence class of functions $v^\ep =\sum a_\mu^\ep e^{i\phi_\mu/\ep} \in C^\infty(\ol{\Omega};\C^N)$ such that, for $\ep \in (0,\ep_0]$, $0<\ep_0<1$,
\begin{itemize}
\item[a)] $|h^\ep - v^\ep_{|t=0}| \leq C_1 \ep^{\frac{1}{2}}$, uniformly in $\ol{X^o}$;
\item[b)] $|L(t,x,\partial) v^\ep | \leq C_2 \ep^{\frac{1}{2}}$, uniformly in $\ol{\Omega}$.
\end{itemize}
\end{proposition}

For every fixed $\ep \in \R_+$, one can apply the classical existence and uniqueness results based on the energy integral method \cite{R,H3} that gives as a byproduct the $L^2$ convergence of the complex geometric optics solution to the exact solution.

\begin{corollary}
\label{conv}
Let $u^\ep,v^\ep \in C^\infty(\ol{\Omega})$ be the exact solution and any representative of the equivalence class in proposition \eref{main-linear}, respectively, then
\begin{equation*}
\sup_{0\leq t \leq T} \|u^\ep(t) - v^\ep(t)\|_{L^2(X^t)} \leq C \ep^{\frac{1}{2}}, \quad \text{for $\ep \in (0,\ep_0]$}.
\end{equation*}
\end{corollary}

In view of the specific construction of the approximate solution, which is based of the smooth reference manifold $R$, it is expected that the latter $L^2$ estimate can be refined to an $L^\infty$ estimate by means of linear conormal estimates; for instance, the case of a single wave $(m=1)$ with a codimension one reference manifold is the analogous of the case considered by Alterman and Rauch in their study of nonlinear geometric optics for short pulses \cite{AR1}. However, we delay the study of the appropriate conormal estimates to the work on generic nonlinear systems.

\section{Analysis of the Matrix $\sigma_{L_0}(t,x,d\phi)$ with $\phi$ a Complex Phase}
\label{sigma}

As a preliminary analysis we study the kernel and the range of the matrix $\sigma_{L_0}(t,x,d\phi)$ for $\phi \in C^\infty(\ol{\Omega};\C)$ being a \emph{single} complex phase. 

Although we know everything about $\sigma_{L_0}(t,x,d\varphi)$ for real-valued $\varphi$, the corresponding results for a complex-valued phase $\phi$ do not follow directly from the assumptions. On the other hand, we note that $|d\chi|$, $\chi = \mathrm{Im} \phi$, is small near the zero level set $R= \{(t,x)\in \Omega;  \chi(t,x)=0\}$ since $\chi$ restricted to any curve $\gamma \subset \ol{\Omega}$ has a minimum in $\gamma \cap R$ when $\gamma$ is transversal to $R$ and it is constant if $\gamma \subset R$; hence it is natural to regard $id\chi$ as a perturbation in the expression $d\phi = d\varphi + id\chi$. 

For any function $f \in C^\infty(\dot{\R}^d;{\bf E})$ taking values in a generic finite-dimensional vector space ${\bf E}$ and for any integer $n \in \N$, let us set
\begin{equation}
\label{cd}
\widetilde{f}^{(n)}(\zeta) = \sum_{|\alpha|\leq n} \frac{i^{|\alpha|}}{\alpha!} \partial^\alpha_\xi f(\xi) \eta^\alpha, \qquad \zeta = \xi + i \eta \in \C^d, \ \xi \not=0.
\end{equation}
We see that $\widetilde{f}^{(n)}$ is of class $C^\infty$ in the complex domain $\dot{\R}^d + i \R^d \subset \C^d$.

\begin{proposition}
\label{prod-ext}
Let $f_i \in C^\infty(\dot{\R}^d;{\bf E}_i)$, $i=1,2,3$, and let us assume that there exists a composition law $:{\bf E}_1 \times {\bf E}_2 \ni (f_1,f_2) \mapsto f_3 = f_1f_2 \in {\bf E}_3$ such that the Leibniz's rule $\partial_{\xi_i} (f_1 f_2) = (\partial_{\xi_i} f_1)f_2 + f_1(\partial_{\xi_i} f_2)$ holds; then, for every $n\in\N$,
\begin{equation*}
 \widetilde{f}_1^{(n)}(\zeta) \widetilde{f}_2^{(n)}(\zeta) = \widetilde{f}_3^{(n)}(\zeta) + \rho^{(n)}(\xi,\eta),
 \end{equation*}
 with $\rho^{(n)}(\xi,\eta) = \sum \frac{i^{|\alpha+\beta|}}{\alpha!\beta!} \partial_\xi^\alpha f_1(\xi) \partial^\beta_\xi f_2(\xi) \eta^{\alpha + \beta}=O(|\eta|^{n+1})$, the sum being over $\alpha,\beta$ with $n < |\alpha + \beta|\leq 2n$.
 \end{proposition}
 
\begin{proof}
By applying the Leibniz's formula we have
\begin{equation*}
\widetilde{f}_3^{(n)}(\zeta) = \sum_{|\gamma|\leq n} \frac{i^{|\gamma|}}{\gamma!} \partial_\xi^\gamma \big(f_1(\xi)f_2(\xi)\big) \eta^\gamma = \sum_{|\gamma|\leq n} \frac{(i\eta)^\gamma}{\gamma!} \sum_{\alpha+\beta=\gamma} \frac{\gamma!}{\alpha!\beta!} \partial_\xi^\alpha f_1(\xi) \partial^\beta_\xi f_2(\xi),
\end{equation*}
whereas,
\begin{align*}
\widetilde{f}_1^{(n)}(\zeta) \widetilde{f}_2^{(n)}(\zeta) &= \sum_{|\alpha|\leq n}\sum_{|\beta|\leq n} \frac{i^{|\alpha+\beta|}}{\alpha!\beta!} \partial_\xi^\alpha f_1(\xi) \partial^\beta_\xi f_2(\xi) \eta^{\alpha+\beta}\\
&= \sum_{|\alpha +\beta|\leq n} \frac{i^{|\alpha+\beta|}}{\alpha!\beta!} \partial_\xi^\alpha f_1(\xi) \partial^\beta_\xi f_2(\xi) \eta^{\alpha+\beta} + \rho^{(n)}(\xi,\eta) \\
&= \widetilde{f}_3^{(n)}(\zeta) + \rho^{(n)}(\xi,\eta).
\end{align*}   
\end{proof}

\begin{remark}
\label{anche}
Clearly the definition of $\widetilde{f}^{(n)}$ and the corresponding property for composition laws still hold if $f$ depends also on $(t,x) \in \ol{\Omega}$.
\end{remark}

We now apply this simple result to the matrices $A(t,x,\xi)$ and $\pi_l(t,x,\xi)$ defined in section \ref{results}, the composition law being the matrix multiplication.

\begin{proposition}
\label{ext-eigenvalue}
For every $n\in\N$,
\begin{itemize}
\item[a)] the exact identity $I = \sum_l \widetilde{\pi}_l^{(n)}$ holds;
\item[b)] the following identities hold modulo $O(|\eta|^{n+1})$, 
\begin{equation*}
\widetilde{\pi}_l^{(n)}\widetilde{\pi}_{l'}^{(n)} = \delta_{ll'}\widetilde{\pi}_l^{(n)}, \quad \widetilde{A}^{(n)} = \sum_l \widetilde{\lambda}^{(n)}_l \widetilde{\pi}_l^{(n)},\quad \widetilde{A}^{(n)} \widetilde{\pi}_l^{(n)} = \widetilde{\lambda}_l^{(n)} \widetilde{\pi}_l^{(n)};
\end{equation*}
\item[c)] there is a unitary matrix $\mathrm{U}(t,x,\xi)$ such that, modulo $O(|\eta|^{n+1})$,
\begin{equation*}
I = \widetilde{\mathrm{U}}^{(n)} \widetilde{\mathrm{V}}^{(n)} = \widetilde{\mathrm{V}}^{(n)} \widetilde{\mathrm{U}}^{(n)}, \qquad \widetilde{\mathrm{U}}^{(n)} \widetilde{A}^{(n)} \widetilde{\mathrm{V}}^{(n)} = \mathrm{diag}(\widetilde{\lambda}^{(n)}_1, \ldots, \widetilde{\lambda}^{(n)}_N),
\end{equation*}
where $\mathrm{V} = \mathrm{U}^*$ and $\mathrm{diag}$ denotes the diagonal matrix, the eigenvalues $\widetilde{\lambda}^{(n)}_l$ being counted with their multiplicity.
\end{itemize}
\end{proposition}

\begin{proof}
We know that $I = \sum_l \pi_l$, $\pi_l\pi_{l'} =\delta_{ll'}\pi_l$, $A = \sum_l \lambda_l \pi_l$ and $A\pi_l = \lambda_l \pi_l$, then we have also $0 = \sum_l \partial_\xi^\alpha \pi_l$ for every multi-index $\alpha$ with $|\alpha| \geq 1$. The latter implies
\begin{equation*}
\sum_l \widetilde{\pi}^{(n)}_l(t,x,\zeta) = \sum_{|\alpha|\leq n} \frac{i^{|\alpha|}}{\alpha!} \sum_l \partial^\alpha_\xi \pi_l(t,x,\xi) \eta^\alpha = \sum_l \pi_l(t,x,\xi) = I,
\end{equation*}
which is the exact identity in a). The relations in b) follow directly form proposition \ref{prod-ext}, for instance the first one reads
\begin{equation*}
\widetilde{\pi}_l^{(n)}\widetilde{\pi}_{l'}^{(n)} = \widetilde{(\pi_l \pi_l')}^{(n)} + O(|\eta|^{n+1}) = \delta_{ll'} \widetilde{\pi}_l^{(n)} + O(|\eta|^{n+1}),
\end{equation*}
and analogously for the others. As for c), we know that $A$ is Hermitian, hence, there exists a unitary matrix $\mathrm{U}(t,x,\xi)$ such that $\mathrm{U}A\mathrm{U}^* = \mathrm{diag}(\lambda_1,\ldots,\lambda_N)$ where the eigenvalues are counted with their multiplicity. Then, the claimed identities follows from proposition \ref{prod-ext}.
\end{proof}

\begin{corollary}
\label{ext-ps}
For every integer $n>0$ and complex phase $\phi$,
\begin{gather*}
-i\sigma_{L_0}(t,x,d\phi) = \sum \big(\partial_t \phi + \widetilde{\lambda}^{(n)}_l(t,x,d_x\phi)\big) \widetilde{\pi}^{(n)}_l(t,x,d_x\phi) + O(|d_x\chi|^{n+1}),\\
\begin{aligned}
\widetilde{\mathrm{U}}^{(n)} (-i\sigma_{L_0}) \widetilde{\mathrm{V}}^{(n)} =  \mathrm{diag}\big(\partial_t \phi + \widetilde{\lambda}^{(n)}_1(t,x,d_x\phi), \ldots,\partial_t \phi + \widetilde{\lambda}^{(n)}_N&(t,x,d_x\phi)\big) \\
&+ O(|d_x\chi|^{n+1}).
\end{aligned}
\end{gather*}
\end{corollary}

\begin{proof}
Since $\partial_\xi^\alpha A(t,x,\xi)=0$ for $|\alpha| > 1$ we have $\widetilde{A}^{(n)}(t,x,\zeta) = A(t,x,\zeta)$ and, by definition \ref{complex_phase}, $d\phi(t,x) \in \dot{\R}^d+i\R^d$; thus, $-i\sigma_{L_0}(t,x,d\phi) = \partial_t \phi I + \widetilde{A}^{(n)}(t,x,d_x\phi)$ for every $n\geq 1$. The claim then follows from proposition \ref{ext-eigenvalue}.
\end{proof}

Corollary \ref{ext-ps} gives enough informations on the kernel and the range of the matrix $\sigma_{L_0}(t,x,d\phi)$ modulo the $O(|d_x\chi|^{n+1})$ remainder. In particular, we see that the kernel is non-trivial if we pick $\phi$ such that 
\begin{equation*}
D^{(n)}_l(t,x,d\phi) = \partial_t \phi + \widetilde{\lambda}_l^{(n)}(t,x,d_x\phi),
\end{equation*}
is $O(|d_x\chi|^{n+1})$ for some $l$, that is, if we find an approximate solution of the complex eikonal equation
\begin{equation*}
D_l^{(n)}\big(t,x,d\phi(t,x)\big)=0,
\end{equation*}
within a fixed order of accuracy.

\section{Approximate Solution of the Complex Eikonal Equation}
\label{gcee}

We will now determine the approximate solution $\phi \in C^{\infty}(\ol{\Omega};\C)$ of the Cauchy problem 
\begin{equation*}
\left\{\begin{aligned}
&\partial_t \phi + \widetilde{\lambda}^{(n)}(t,x,d_x\phi) =0, \quad &(t,x)\in \ol{\Omega} \\
&\phi_{|t=0}(x) = \psi(x), &x \in \ol{X^o},
\end{aligned}\right.
\end{equation*}
with $\lambda(t,x,\xi) \in C^\infty(\ol{\Omega}\times \dot{\R}^d)$, homogeneous of degree 1 in $\xi$, and $\psi \in C^{\infty}(\ol{X^o};\C) $. The analogous of assumptions \ref{Rzero} and \ref{ref-man} with $m=1$ and $ f(t,x,\tau,\xi)=\tau + \lambda(t,x,\xi)$ are supposed to be true.

According to assumption \ref{ref-man}, there is an isotropic submanifold $\Lambda$ which is determined by the flow-out along the Hamiltonian vector field $H_f$ in $T^*\R^{1+d}$ of the isotropic submanifold given over the hyperplane $\{t=0\}$ by the graph of $d\mathrm{Re}\psi(x)$ for $x \in R^o$; the projection of $\Lambda$ onto $\ol{\Omega}$ amounts to the submanifold $R$ which is supposed to be transverse to the boundary $X^T$ of the domain $\ol{\Omega}$. If the mapping $F:[0,T]\times R^o \to \ol{\Omega}$ is the projection on $\ol{\Omega}$ of the Hamiltonian flow, then $R = F([0,T]\times R^o)$ and we have the following proposition.

\begin{proposition}
\label{refman-properties}
If assumption \ref{ref-man} is satisfied,
\begin{itemize}
\item[a)] for every $t \in [0,T]$, $R^t = X^t \cap R$ is a smooth submanifold of $X^t$ and $R^t \cong R^o$;
\item[b)] $R$ is fibered over $R^o$, the fibers being the integral curves of the smooth vector field $V(t,x)= \partial_t + d_\xi \lambda(t,x)$ restricted to $R$.
\end{itemize}
\end{proposition} 

\begin{proof}
According to assumption \ref{ref-man}, the Hamiltonian flow amounts to a diffeomorphism $:[0,T] \times R^o \to \Lambda \cap \{(t,x,\tau,\xi); (t,x) \in \ol{\Omega}\}$ and $\Lambda \cap \{(t,x,\tau,\xi); (t,x) \in \ol{\Omega} \}$ is diffeomorphic to $R$, thus, $F$ is an embedding of $[0,T]\times R^o$ into $\ol{\Omega}$ and $F^{-1}$ is well-defined on $R$; specifically, $F^{-1}(t,x) =(t,x^o)$ with $x = x(t,x^o)$ being the projection of the Hamiltonian orbit issuing from $(x^o, d\psi(x^0))$ ($d\psi(x)$ is real when $x \in R^o$). 

a) The mapping $F^t = F(t,\cdot ) :R^o \to X^t$ given by $x^o \mapsto (t,x(t,x^o))=F(t,x^o)$ is an embedding of $R^o$ in $\ol{\Omega}$ for every $t \in [0,T]$, therefore $R^t =F^t(R^o)$ is a smooth manifold without boundary diffeomorphic to $R^o$.

b) On fixing some local coordinates $r$ in an open neighbourhood $\mathcal{R}^o \subseteq R^o$ we have natural local coordinates in the neighbourhood $\mathcal{R}=F([0,T] \times \mathcal{R}^o) \subseteq R$ given by $(t,x) \mapsto (t, r)$, with $r$ defined as the coordinates in $\mathcal{R}^o$ of the point $x^o$ such that $x=x(t,x^o)$. Moreover, we have the projection $\Pi : R \ni (t,x) \mapsto x^0 \in R^o$, where, again, $x$ and $x^o$ are related by $x = x(t,x^o)$; in coordinates $\Pi$ amounts to the projection on the second factor $(t,r) \mapsto r$. The fibers of $\Pi$ are $\Pi^{-1}(x^o) = \{(t,x(t,x^o)) \in R ; t\in [0,T]\}$, that is the integral curve of the vector field $V$.
\end{proof}

We shall now extend the fibered coordinates $(t,r)$ to coordinates $(t,r,s)$ in a neighbourhood $\mathcal{O} \subseteq \ol{\Omega}$. With this aim, we make use of the Euclidean metric $\langle \cdot,\cdot  \rangle$ in order to define the normal space of $R^t$ in $X^t$ at the point $(t,x) \in R^t$, namely,
\begin{equation*}
N_{(t,x)}R^t = \{ \delta x \in T_{(t,x)}X^t; \langle \delta r,\delta x \rangle=0, \forall \delta r \in T_{(t,x)}R^t\},
\end{equation*}
and the disjoint union for $(t,x)\in R^t$ of $N_{(t,x)}R^t$ defines the normal bundle $NR^t$. Then, the disjoint union
\begin{equation*}
\mathcal{T} = \bigcup_{0\leq t \leq T}  TX^t_{|R^t}, \qquad \mathcal{N} = \bigcup_{0 \leq t \leq T} NR^t,
\end{equation*}
are smooth vector bundles over $R$ locally isomorphic to $R \times \R^d$ and $R \times \R^{d_2}$, respectively. To see this, let $e^o \in N_{x^o}R^o$, $\delta r^o \in T_{x^o}R^o$ and $\delta r =DF^t \delta r^o \in T_{F^t(x^o)}R^t$ with $DF^t$ be the Jacobian matrix of the diffeomorphism $F^t : R^o \to R^t \subset X^t$ for $t\in[0,T]$; then,
\begin{equation*}
0 = \langle e^o,\delta r^o \rangle = \langle e^o, (DF^t)^{-1}\delta r \rangle = \langle {}^t(DF^t)^{-1}e^o, \delta r \rangle,
\end{equation*}
and $^t(DF^t)^{-1}$ defines a local bundle isomorphism $:NR^o \to NR^t$ for all $t \in [0,T]$. Therefore, one can extend any vector bundle chart on $NR^o$ to $NR^t$ smoothly for all $t \in [0,T]$, thus, obtaining a vector bundle chart for $\mathcal{N}$ of the form $[0,T]\times \mathcal{R}^o \times \R^{d_2}$; as for $\mathcal{T}$, we can construct a vector bundle chart on noting that $T_{(t,x)}X^t_{|R^t} = T_{(t,x)}R^t \oplus N_{(t,x)}R^t$ and on applying the same argument to the relation $\delta r = DF^t \delta r^o$. Furthermore, $\mathcal{N}$ is a subbundle of $\mathcal{T}$ and let $\Gamma$ be the projector $:\mathcal{T} \to \mathcal{N}$; when, evaluated in a point $(t,x) \in R$, $\Gamma(t,x)$ amounts the projector $:T_{(t,x)}X^t \to N_{(t,x)}R^t$. An orthogonal frame for $\mathcal{N}$ can be conveniently obtained by solving the Cauchy problem 
\begin{equation*}
\left\{
\begin{aligned}
& e_i(t,r) \in C^\infty(R;\mathcal{N}),\\
& \big[\Gamma \partial_t \Gamma + (I-\Gamma) \partial_t (I-\Gamma)\big]e_i = \sum_{j=1}^{d_2} c_{ij} \Gamma e_j,\\
& e_{i|t=0}(r) = e^o_i(r) \in C^\infty(R^o;NR^o),
\end{aligned}
\right.
\end{equation*}
with orthogonal initial data, namely, $\langle e^o_i, e^o_j \rangle=\delta_{ij}$, and with $c_{ij} \in C^\infty(R;\R)$, $c_{ij} + c_{ji}=0$. Since $T_{(t,x)}X^t =\R^d$, $N_{(t,x)}R^t  \subset \R^d$, we have $\Gamma(t,x) \in \mathrm{End}(\R^d)$ and $e_i \in \R^d$; indeed, that amounts to a system of  linear ordinary differential equations in $\R^d$ for which the solution exists in $[0,T]$ and it is unique. On the other hand, we see that, if $\{e_i\}_i$ is a solution, also $\{ \Gamma e_i \}_i$ does, hence, it should be $e_i = \Gamma e_i$ or equivalently, $e_i \in C^\infty(R;\mathcal{N})$ as required. Finally, such a solution defines an orthogonal frame for $\mathcal{N}$. The orthogonality can be proved on considering the matrix $E =(E_{ij})$ whose elements are just the scalar products $E_{ij}=\langle e_{i}, e_{j} \rangle$. If $e_i$ is a solution, then $\partial_te_i = [2\Gamma (\partial_t\Gamma) - \partial_t\Gamma]e_i + \sum_j c_{ij}\Gamma e_j = [\partial_t\Gamma,\Gamma]e_i + \sum_{j} c_{ij}e_j$ and, on noting that the commutator is anti-symmetric, i.e., ${}^t[\partial_t\Gamma,\Gamma]=-[\partial_t\Gamma,\Gamma]$ that follows from ${}^t\Gamma =\Gamma$, one finds
\begin{equation*}
\partial_t E = CE + E\ {}^tC, \qquad E_{|{t=0}} =I,
\end{equation*}   
where $C=(c_{ij})$ is anti-symmetric. By direct substitution we see that $E(t,r)=I$ is the unique solution proving the orthogonality. One should note that the construction of the basis $\{e_i(t,r)\}$ does not depend on the coordinates $r$ on $R^o$, i.e., the projector $\Gamma(t,r)$ and vectors $e_i(t,r)$ are invariant under a coordinate change in $R^o$.

We can now define coordinates $[0,T] \times \mathcal{O}_r\times \mathcal{O}_s \ni (t,r,s) \mapsto (t, x) \in \ol{\Omega}$ by
\begin{equation}
\label{coordinates}
x = x(t,r,s) = x(t,r) + \sum e_j(t,r) s_j,
\end{equation}
for suitably small neighbourhoods $\mathcal{O}_r \subseteq \R^{d_1}$ and $\mathcal{O}_s \subseteq \R^{d_2}$. We see that the differential of the map \eref{coordinates} evaluated at $s=0$ amounts to
\begin{equation*}
\left( \frac{\partial x(t,r)}{\partial t}, \frac{\partial x(t,r)}{\partial r_1},\ldots, \frac{\partial x(t,r)}{\partial r_{d_1}}, e_1(t,r),\ldots,e_{d_2}(t,r) \right),
\end{equation*}
which has a non-zero determinant as the columns are linearly independent. Therefore, the map \eref{coordinates} is a local diffeomorphism in a neighbourhood $\mathcal{O}$, with $\mathcal{R} \subset \mathcal{O} \subset \ol{\Omega}$ and with boundary $\partial\mathcal{O} = \mathcal{O}^o \cup \mathcal{O}^T$, where $\mathcal{O}^t = \mathcal{O} \cap X^t$ are open in $X^t$. These are the coordinates used in the formulation of lemma \ref{maslov-estimate2-small}.

\begin{remark}
\label{beam-coordinates}
For the degenerate case $R^o = \{x^o\}$, the coordinates $s$ are simply given by $s = x-x(t,x^o)$. However, one can also make use of the Frenet frame associated to the curve $R$, \cite{Kr, KrBe}. This differs from the construction of local coordinates described above in the fact that our normal vectors $e_i$ lie in $t= \text{constant}$ hyperplanes.  
\end{remark}

We will make use of such coordinates to find an asymptotic solution of the Cauchy problem for the complex eikonal equation. Specifically, let us set
\begin{equation*}
\phi(t,r,s) = \sum_{|\alpha| \leq n} \frac{1}{\alpha!} \phi_\alpha(t,r)s^\alpha,
\end{equation*}
the sum being over the multi-index $\alpha \in \N^{d_2}$; we want to choose the coefficients $\phi_\alpha(t,r)$ so that 
\begin{itemize}
\item the manifold $R=F([0,T]\times R^o)$ is the reference manifold for $\phi$,
\item the complex eikonal equation is satisfied near $R$ modulo $O(s^{n+1})$, i.e, we have
\begin{equation}
\partial^\alpha_s D^{(n)}\big(t,r,s, d\phi(t,r,s) \big)_{|R} = 0, \quad \text{for}\quad |\alpha| \leq n.
\end{equation}
\end{itemize}
Here, the subscript $|R$ means that the function should be evaluated for $s=0$.

We see that $\phi_0 (t,r) = \phi_{|R}(t,r)$ whereas for $|\alpha| = 1$, $\phi_\alpha$ amounts to $\phi_j(t,r) = \partial_{s_j} \phi_{|R}(t,r)$ and, since $R$ should be the reference manifold of $\phi$, we must have $\phi_0 (t,r) = \varphi_0(t,r) \in \R$, and $\phi_j(t,r) = \partial_{s_j}\varphi_{|R}(t,r) \in \R$, for $j \in \{1,\ldots,d_2\}$. Then, the lowest non-trivial order, for which complex-valued phases are found, is $n=2$, namely,
\begin{subequations}
\label{eikonal}
\begin{equation}
\label{complex-eikonal}
\phi(t,r,s) = \varphi_0(t,r) + \sum_{i=1}^{d_2} \varphi_i(t,r) s_i + \frac{1}{2} \sum_{i,j=1}^{d_2} \phi_{ij}(t,r) s_is_j,
\end{equation}
and we shall make use of notations
\begin{align}
\label{real-eikonal}
&\mathrm{Re}\phi(t,r,s)= \varphi(t,r,s) = \varphi_0(t,r) + \sum_{i=1}^{d_2} \varphi_i(t,r) s_i + \frac{1}{2} \sum_{i,j=1}^{d_2} \varphi_{ij}(t,r) s_is_j,\\
\label{imaginary-eikonal}
&\mathrm{Im}\phi(t,r,s) = \chi(t,r,s) = \frac{1}{2} \sum_{i,j=1}^{d_2} \chi_{ij}(t,r)s_is_j.
\end{align}
\end{subequations} 
The matrix $(\chi_{ij})$ should be positive definite. As a direct consequence we have,
\begin{gather*}
\frac{\partial \phi}{\partial r_k}\Big|_R = \frac{\partial \varphi_0}{\partial r_k}, \quad \frac{\partial \phi}{\partial s_j} \Big|_R = \varphi_{j}, \quad \frac{\partial^2 \phi}{\partial r_k\partial r_h}\Big|_R = \frac{\partial^2 \varphi_0}{\partial r_k \partial r_h},\\
\frac{\partial^2 \phi}{\partial r_k \partial s_j}\Big|_R = \frac{\partial \varphi_j}{\partial r_k}, \quad \frac{\partial^2 \phi}{\partial s_i \partial s_j} \Big|_R = \phi_{ij}.
\end{gather*}
From now on we consider only the lowest non-trivial order and we drop the index $n=2$. First, we note that, for every $t\in [0,T]$, the function 
\begin{equation*}
\widetilde{\lambda}(t,x,\zeta) = \lambda(t,x,\xi ) + i \sum_{i=1}^{d} \frac{\partial \lambda(t,x,\xi)}{\partial \xi_i} \eta_i - \frac{1}{2} \sum_{i,j=1}^d \frac{\partial^2\lambda(t,x,\xi)}{\partial \xi_i \partial \xi_j} \eta_i \eta_j
\end{equation*}
is invariantly defined if $(x,\xi,\eta)\in T^*X^t \oplus T^*X^t$, where $\oplus$ denotes the Whitney sum of vector bundles. Thus, we can readily write $\widetilde{\lambda}(t,x,d_x\phi)$ in terms of coordinates $(t,r,s)$, namely,
\begin{multline*}
\widetilde{\lambda}(t,x,d\phi(t,x)) = \Lambda(t,r,s,d_r\varphi, d_s\varphi) + i \sum_{k} \frac{\partial \chi}{\partial r_k}\frac{\partial \Lambda}{\partial \rho_k} (t,r,s,d_r\varphi, d_s\varphi)  \\
+ i \sum_j \frac{\partial \chi}{\partial s_j} \frac{\partial \Lambda}{\partial \sigma_j}(t,r,s,d_r\varphi, d_s\varphi) -\frac{1}{2} \sum_{h,k} \frac{\partial \chi}{\partial r_k}\frac{\partial \chi}{\partial r_h} \frac{\partial^2 \Lambda}{\partial \rho_h\partial \rho_k} (t,r,s,d_r\varphi, d_s\varphi) \\
- \sum_{k,j} \frac{\partial \chi}{\partial r_k} \frac{\partial \chi}{\partial s_j} \frac{\partial^2 \Lambda}{\partial \rho_k \partial \sigma_j}(t,r,s,d_r\varphi, d_s\varphi) - \frac{1}{2} \sum_{i,j}  \frac{\partial \chi}{\partial s_i}\frac{\partial \chi}{\partial s_j} \frac{\partial^2 \Lambda}{\partial \sigma_i \partial \sigma_j} (t,r,s,d_r\varphi, d_s\varphi),
\end{multline*}
with $(\rho, \sigma)$ are the coordinates dual to $(r,s)$ in $T^*X^t$ and $\Lambda(t,r,s,\rho,\sigma)$ is the pull-back of $\lambda(t,x,\xi)$ under the change of coordinates $(t,r,s,\rho,\sigma) \mapsto (t,x,\xi)$.

We readily find that the  equation $D_{|R}(t,r) =0$ implies 
\begin{subequations}
\label{inutili}
\begin{equation}
\label{hj}
\partial_t \varphi_0(t,r) + \Lambda(t,r,s,d_r\varphi, d_s\varphi)_{|R}=0,
\end{equation}
where $d_r\varphi_{|R} = d_r\varphi_0$ and $d_s\varphi_{|R}=\sum \varphi_ids_i$. Analogously, from the real and imaginary parts of $\partial_{s_i} D_{|R}(t,r) =0 $ we get
\begin{equation}
\label{null}
\partial_t\varphi_i + \partial_{s_i}\Lambda + \sum_k \partial_{\rho_k}\Lambda \partial_{r_k}\varphi_i + \sum_j \partial_{\sigma_j}\Lambda \varphi_{ij} =0,\quad \sum_j \partial_{\sigma_j}\Lambda \chi_{ij}  =0,
\end{equation}
\end{subequations}
and, finally, from the real and imaginary parts of $\partial_{s_is_j}D_{|R}(t,r)$ we have
\begin{gather*}
\partial_t \varphi_{ij} + A_{ij} + \sum_l \varphi_{il} B_{lj} + \sum_l {}^tB_{il} \varphi_{lj} + \sum_{k,l} \varphi_{ik} C_{kl}\varphi_{lj} - \sum_{k,l} \chi_{ik}C_{kl}\chi_{lj} = 0,\\
\partial_t \chi_{ij} + \sum_l \chi_{il} B_{lj} + \sum_l {}^tB_{il} \chi_{lj} + \sum_{k,l} \varphi_{ik}C_{kl}\chi_{lj} + \sum_{k,l}\chi_{ik}C_{kl}\varphi_{lj} =0,
\end{gather*}
or, equivalently,
\begin{equation}
\label{riccati}
\partial_t \phi_{ij} + A_{ij} + \sum_l \phi_{il} B_{lj} + \sum_l {}^tB_{il} \phi_{lj} + \sum_{k,l} \phi_{ik} C_{kl}\phi_{lj} =0,
\end{equation}
where ${}^tB_{ij} = B_{ji}$ is the transpose matrix and the coefficients are
\begin{gather*}
A_{ij}(t,r) = \frac{\partial^2\Lambda}{\partial s_i \partial s_j} + \sum_k \frac{\partial^2\Lambda}{\partial s_i \rho_k}\frac{\partial \varphi_j}{\partial r_k} + \sum_{k} \frac{\partial \varphi_i}{\partial r_k} \frac{\partial^2\Lambda}{\partial \rho_k\partial s_j} + \sum_{h,k} \frac{\partial^2\Lambda}{\partial\rho_h \partial \rho_k} \frac{\partial \varphi_i}{\partial r_h} \frac{\partial \varphi_j}{\partial r_k},\\
B_{ij}(t,r) = \frac{\partial^2\Lambda}{\partial \sigma_i\partial s_j} + \sum_k \frac{\partial^2\Lambda}{\partial \sigma_i\partial \rho_k}\frac{\partial \varphi_j}{\partial r_k},\qquad C_{ij}(t,r) = \frac{\partial^{2}\Lambda}{\partial \sigma_i \partial \sigma_j}.
\end{gather*}
In all the foregoing expressions, the arguments the the derivatives of $\Lambda$ are evaluated at $(t,r,0, d_r\varphi_0, \sum\varphi_ids_i)$.

\begin{remark}
\label{riccati-rem}
Equation \eref{riccati} constitutes a family of matrix Riccati equations and its unique solution exists in $C^\infty([0,T_1])$, $0 < T_1 \leq T$. In addition, we require that $(\chi_{ij}(0,r))$ is symmetric and positive definite and it is not difficult to show that $(\chi_{ij})$ keeps symmetric and positive definite for $t \in [0,T_1]$. Therefore, at least after redefining $T$, the solution of \eref{riccati} exists in $[0,T]$ with $(\chi_{ij})$ symmetric and positive definite.
\end{remark}

\begin{remark}
\label{BT-equations}
For the degenerate case $R^o=\{x^o\}$ the coefficients take the form
\begin{equation*}
A_{ij}(t,r) = \frac{\partial^2\Lambda}{\partial s_i \partial s_j}, \quad B_{ij}(t,r) = \frac{\partial^2\Lambda}{\partial \sigma_i\partial s_j}, \quad C_{ij}(t,r) = \frac{\partial^{2}\Lambda}{\partial \sigma_i \partial \sigma_j},
\end{equation*}
which is the form known as beam tracing equations in the applied literature [4-7].
\end{remark}

In view of assumption \ref{ref-man}, the isotropic manifold $\Lambda$ over the relatively open set $\mathcal{O} \subseteq \ol{\Omega}$ can be parametrized by coordinates $(t,r)$, that is, points on $\Lambda \cap \{(t,x,\tau,\xi); (t,x)\in \mathcal{O} \}$ are of the form $(x(t,r),\xi(t,r))$ with $(t,r) \in [0,T]\times \mathcal{R}^o$.

\begin{proposition}
\label{reduction-to-riccati}
Let assumption \ref{ref-man} be verified. If we set $\varphi_0(t,r) = \psi(x^o(r))$ and $d_{(r,s)}\varphi_{|R}(t,r) = \xi(t,r)$, equations \eref{inutili} are identically satisfied. In addition, if $\phi_{ij}(t,r)$ solves the Riccati equation \eref{riccati} for $t\in[0,T]$ with initial condition $\partial_{s_is_j} \psi_{|R^o}(r)$, where $\psi$ is written in terms of the coordinates $(r,s)$ near $R^o$, then the complex phase \eref{eikonal} is a solution modulo $O(|s|^3)$ of the Cauchy problem for the complex eikonal equation, that is, a) $\partial_t\phi + \widetilde{\lambda}(t,x,d_x\phi) = O(|s|^3)$, in $\mathcal{O}$, and b) $\phi(0,r,s) - \psi(r,s) = O(|s|^3)$, in $\mathcal{O}^o$.
\end{proposition}
 
\begin{proof}
The pull back of the vector field $\partial_t + d_\xi\lambda(t,x,d_x\varphi)$ to the local coordinates $(t,r,s)$ reads $\partial_t + d_{(\rho,\sigma)}\Lambda(t,r,s,d_{r}\varphi,d_s\varphi)$, and the orbits of the latter field are just the orbits of the former field pulled back in the local coordinates. We know that the orbits on $R$ of $\partial_t + d_\xi\lambda(t,x(t,r),\xi(t,r))$ in local coordinates are $\{(t,r,0); t\in [0,T], r=\text{const.}\}$, then $\partial_t + d_{(\rho,\sigma)}\Lambda(t,r,s,d_{r}\varphi,d_s\varphi)_{|R} = \partial_t$ and the partial derivatives $\partial_{\rho} \Lambda$, $\partial_{\sigma}\Lambda$ vanishes when evaluated for $s=0$, $(\rho,\sigma)=\xi(t,r)$.

On recalling that $\lambda$ and $\Lambda$ are homogeneous functions of degree one in the dual variables $\xi$ and $(\rho,\sigma)$, respectively, we can apply the Euler formula for which
\begin{equation*}
\partial_t \varphi_0 + \Lambda(t,r,s,d_r\varphi, d_s\varphi)_{|s=0}=\partial_t \varphi_0 + \sum \frac{\partial \Lambda}{\partial \rho_k} \frac{\partial \varphi_0}{\partial r_k} + \sum \frac{\partial \Lambda}{\partial \sigma_j} \varphi_j = \partial_t \varphi_0 =0,
\end{equation*}
which is solved by $\varphi_0(t,r) = \psi_{|R^o}(r)$ as claimed ($\psi_{|R^o}$ is real). Analogously, equation \eref{null} reduces to
\begin{equation*}
\partial_t\varphi_i + \partial_{s_i}\Lambda =0,
\end{equation*}
which is just the $s$-component of the Hamilton's equation $d\xi/dt + d_x\lambda=0$ expressed in the local coordinates, hence, it is identically satisfied. The only remaining equation is the Riccati equation \eref{riccati} which, indeed, is a nonlinear ordinary differential equation. If $\phi_{ij}(t,r)$ is a solution we have $D(t,x,d\phi) = O(|s|^3)$ and a) is proved. Analogously, b) follows from the Taylor expansion of the initial datum $\psi$.
\end{proof}

The vice versa is also valid, that is, any approximate solution of the complex eikonal equation with the same initial datum should have the same reference manifold, determined by the characteristics, and the same Taylor polynomial around it up to the prescribed order; the proof requires the following lemma.

\begin{lemma}
\label{Gronwall-type}
Let $\delta,t_0 \in \R_+$, $f \in C(\ol{\R}_+)$, and let us suppose that a non-negative function $\chi \in C^1([0,t_0];\ol{\R}_+)$ is such that $\chi(0)=0$ and $\chi(t) \leq \delta$ implies $\chi'(t) \leq f(\chi(t))\chi(t)$. Then, $\chi(t)=0$ for all $t\in [0,t_0]$.
\end{lemma}

\begin{proof}
Let $t_* = \sup \{t_1 \in [0,t_0];\ \chi(t)=0\ \text{when}\ t\in [0,t_1]\}$; clearly $t_* \geq 0$. Since $\chi \in C^1$, if $t_* < t_0$, there exists $t_*' > t_*$ such that $\chi(t) \leq \delta$ when $t \in [0,t_*']$. We define $k(t) = \chi(t) \exp\big(-\int_{0}^t f(\chi(t'))dt'\big)$ and find $k'(t) \leq 0$ in $[0,t'_*]$ while $k(0) =0$. This implies $\chi(t)=0$ for $t\in[0,t_*']$ against the definition of $t_*$, hence, the only possibility is $t_*=t_0$.
\end{proof}

\begin{proposition}
\label{viceversa}
Let assumption \ref{ref-man} be satisfied and $\phi =\varphi + i \chi \in C^3(\ol{\Omega};\C_+)$ be a complex phase and let $R=\{(t,x) \in \ol{\Omega}; \chi(t,x)=0\}$ be a submanifold, with $(t,r,s)$ generic coordinate having the submanifold property for $R$. If $\phi$ solves, modulo $O(|s|^3)$, the Cauchy problem for the complex eikonal equation with initial value $\psi$ satisfying assumption \ref{Rzero}, then, $R=F([0,T]\times R^o)$ and $\phi$ has the Taylor expansion \eref{eikonal} with coefficients given in proposition \ref{reduction-to-riccati}.
\end{proposition}

\begin{proof}
Since $R\cap X^o=R^o$, we have $\mathrm{dim} R = \mathrm{dim}R^o +1$. 
According to the hypotheses, there are constants $k_1$, $k_2$ such that $|s|^2 k_1 \leq \chi \leq k_2 |s|^2$ at least for $|s|$ small enough. Therefore, from the imaginary part of the complex eikonal equation near $R$, one finds
\begin{equation*}
\partial_t\chi + \langle d_\xi\lambda(t,x,d_x\varphi),d_x\chi \rangle \leq C \chi^{3/2}.
\end{equation*}
We pick an integral line $t \mapsto (t,x(t))$ of the vector field $\partial_t + d_\xi\lambda(t,x,d_x\varphi)$ issuing from a point in $R^o$ and apply lemma \ref{Gronwall-type} to $\chi(t,x(t))$ with the result that $\chi(t,x(t))=0$ implying that the integral lines of $V$ issuing from points in $R^o$ stay inside $R$. By applying the operator $\partial_{x_i}$ to the real part of the complex eikonal equation we have
\begin{equation*}
\partial_t \partial_{x_i} \varphi + \partial_{x_i}\lambda(t,x,d_x\varphi) + \sum_j \frac{\partial\lambda}{\partial \xi_j} \frac{\partial^2\varphi}{\partial x_i\partial x_j} =O(|d_x\chi|).
\end{equation*}
At the beginning of section \ref{sigma}, we have already argued that $d_x\chi_{|R}=0$, hence, the latter equation implies
\begin{equation*}
\mathcal{L}_V d_x \varphi_{|R} + d_x \lambda(t,x,d_x\varphi)_{|R}=0,
\end{equation*}
with $\mathcal{L}_V$ the Lie derivative along $V$. Therefore, along the integral lines of $V$ we have the Hamilton's equations
\begin{equation*}
dx/dt = d_\xi\lambda(t,x,\xi), \qquad d\xi/dt = -d_x\lambda(t,x,\xi),
\end{equation*}
where $\xi = d_x\varphi$ and $d\xi/dt = \sum_i (\mathcal{L}_V \partial_{x_i}\varphi)dx_i$. In view of assumption \ref{ref-man}, the solution $x(t,x^o)$, with initial position $x^o \in R^o$, exists for $t\in[0,T]$ and spans a manifold of dimension $1+\dim R^o = \dim R$ that should lie within $R$; we conclude that the whole reference manifold $R$ is spanned by solutions of the Hamilton's equations and this justifies our construction. Finally, a straightforward application of the Taylor formula applied to $D(t,x,d\phi)$ shows that the exact solution agrees with the asymptotic one \eref{eikonal} modulo $O(|s|^3)$ in the neighbourhood $\mathcal{O}\subset \ol{\Omega}$ where local coordinates $(t,r,s)$ are defined.
\end{proof}

Proposition \ref{reduction-to-riccati} allows us to achieve an asymptotic solution to the complex eikonal equation in a neighbourhood $\mathcal{O} \supseteq \mathcal{R}$ where local coordinates $(t,r,s)$ are defined. The relatively open set $\mathcal{R} \subset R$ is found in the form $\mathcal{R} = F([0,T] \times \mathcal{R}^o)$ with $\mathcal{R}^o \subseteq R^o$ being the domain of definition for the local coordinates $r$. Therefore, one should address how to glue two solution based on coordinate defined in two different sets $\mathcal{R}_1^o$, $\mathcal{R}^o_2$, with $\mathcal{R}^o_1\cap\mathcal{R}^o_2 \not = \emptyset$. This is straightforward since the normal vectors $e_i$ and, thus, the coordinates $s$, do not depend on the choice of coordinates $r$ on $R^o$. However, they behave non-trivially under a change of normal vectors $e_i^o$.

We conclude this section with an important lemma which means that a single complex phase $\phi$ can solve the complex eikonal equation corresponding \emph{only} to a specific eigenvalue $\lambda$ in an open subset $\mathcal{O} \subseteq \ol{\Omega}$.

\begin{lemma}
\label{coherence}
If the complex phase $\phi \in C^3(\mathcal{O})$ is a solution of $\partial_t \phi + \widetilde{\lambda}(t,x,d_x\phi) = O(|s|^3)$ with $\lambda(t,x,\xi)$ being an eigenvalue of $A(t,x,\xi)$, for any other eigenvalue $\lambda_l \not = \lambda$ there are constants $c_l >0$ and $s_l>0$ such that
\begin{equation*}
|\partial_t \phi + \widetilde{\lambda}_l(t,x,d_x\phi)| \geq c_l, \quad \text{for}\quad |s| \leq s_l.
\end{equation*}
\end{lemma}

\begin{proof}
Let us write
\begin{align*}
|\partial_t \phi + \widetilde{\lambda}_l(t,x,d_x\phi)| &= \big|\big(\partial_t \phi + \widetilde{\lambda}(t,s,d_x\phi)\big) + \big(\widetilde{\lambda}_l(t,x,d_x\phi) - \widetilde{\lambda}(t,s,d_x\phi)\big)\big| \\
&\geq \Big|\big|\partial_t \phi + \widetilde{\lambda}(t,s,d_x\phi) \big| - \big|\widetilde{\lambda}_l(t,x,d_x\phi) - \widetilde{\lambda}(t,s,d_x\phi) \big| \Big|.
\end{align*}
We have $|\partial_t \phi + \widetilde{\lambda}(t,x,d_x\phi)| \leq C_1|s|^3$, and, with $\chi_i = (d_x\chi)_i$,
\begin{multline*}
\big|\widetilde{\lambda}_l(t,x,d_x\phi) - \widetilde{\lambda}(t,s,d_x\phi) \big| \geq  \Big| \big|\lambda_l(t,x,d_x\phi) - \lambda(t,s,d_x\phi)\big| \\
- \big|\frac{1}{2} \sum_{i,j} \frac{\partial^2\lambda}{\partial\xi_i \partial\xi_j} \chi_i \chi_j  - \frac{1}{2} \sum_{i,j} \frac{\partial^2\lambda_l}{\partial\xi_i \partial\xi_j} \chi_i \chi_j\big| \Big|,
\end{multline*}
where $\big|\lambda_l(t,x,d_x\phi) - \lambda(t,s,d_x\phi) \big| \geq C_3>0$ in view of assumption \ref{modes} and 
\begin{equation*}
\Big|\frac{1}{2} \sum_{i,j} \frac{\partial^2\lambda}{\partial\xi_i \partial\xi_j} \chi_i \chi_j  - \frac{1}{2} \sum_{i,j} \frac{\partial^2\lambda_l}{\partial\xi_i \partial\xi_j} \chi_i \chi_j \Big| \leq C_2 |s|^2.
\end{equation*}
Pick $s_* < (C_3/C_2)^{1/2}$, then, for $|s|\leq s_*$,
\begin{equation*}
\big|\lambda_l(t,x,d_x\phi) - \lambda(t,s,d_x\phi)\big| \geq C_3 \geq \big|\frac{1}{2} \sum_{i,j} \frac{\partial^2\lambda}{\partial\xi_i \partial\xi_j} \chi_i \chi_j  - \frac{1}{2} \sum_{i,j} \frac{\partial^2\lambda_l}{\partial\xi_i \partial\xi_j} \chi_i \chi_j\big|, 
\end{equation*}
hence,
\begin{equation*}
\big|\widetilde{\lambda}_l(t,x,d_x\phi) - \widetilde{\lambda}(t,s,d_x\phi) \big| \geq  C_* = C_3-C_2s_*^2>0,\qquad |s| \leq s_*.
\end{equation*}
Analogously, if we pick $s_l < \min \{s_*, (C_*/C_1)^{1/3}\}$, we have $C_1 s_l^3 < C_*$ and, thus, $\big|\partial_t \phi + \widetilde{\lambda}(t,s,d_x\phi) \big| \leq C_*$, and for $|s| \leq s_l$, 
\begin{equation*}
|\partial_t \phi + \widetilde{\lambda}_l(t,x,d_x\phi)| \geq C_* - C_1s_l^3=c_l > 0.
\end{equation*} 
\end{proof}

\begin{remark}
\label{shorter}
A shorter proof of lemma \ref{coherence} is obtained on noting that, if the claim is not true, for every $C>0$ there is $\lambda_l$ such that $|\lambda-\lambda_l| \leq C$ at least in a point on $\Lambda$ against assumption \ref{modes}. The extended proof given above is somewhat more informative.
\end{remark}

\section{The Amplitudes}
\label{amplitudes-linear}

Let $\phi$ be an approximate solution of the complex eikonal equation addressed in section \ref{gcee}, with $\lambda$ a generic eigenvalue of $A$ corresponding to the projector $\pi$; in this section we shall address equations \eref{eq1} and \eref{eq2} with $\phi_\mu = \phi$ and $q=2$; upon writing $a_\mu^{(0)} = a_0$ and $a_\mu^{(1)}=a_1$ for simplicity, we shall study the equations
\begin{align}
\label{eq1b}
&\sigma_{L_0}(t,x,d\phi)a_0(t,x)=O(|s|^3),\\
\label{eq2b}
&\sigma_{L_0}(t,x,d\phi)a_1(t,x) + L(t,x,\partial)a_0(t,x) =O(|s|),
\end{align}
in a neighbourhood of $R=\{(t,x)\in \ol{\Omega}; \mathrm{Im}\phi(t,x)=\chi(t,x)=0\}$ where coordinates $(t,r,s)$ are defined; we recall that, according to proposition \eref{viceversa}, $R$ is the flow out of $R^o=R\cap \{t=0\}$ along the characteristics of $L_0$. 

We have $a_0 = (I-\widetilde{\pi})a_0 + \widetilde{\pi}a_0$, with $\widetilde{\pi}$ the extended projector (with $n=2$), and 
\begin{equation*}
\sigma_{L_0}(t,x,d\phi) \widetilde{\pi} a_0 = i\big(\partial_t\phi + \lambda(t,x,d_x\phi) \big)\widetilde{\pi}a_0 + O(|s|^3) = O(|s|^3),
\end{equation*}
in view of corollary \ref{ext-ps}. Therefore,
\begin{multline*}
\sigma_{L_0}(t,x,d\phi) a_0 = \sigma_{L_0}(t,x,d\phi) (I-\widetilde{\pi})a_0 + O(|s|^3) \\
= \sum_{l;\ \lambda_l \not = \lambda} i \big(\partial_t\phi + \widetilde{\lambda}_l(t,x,d_x\phi) \big) \widetilde{\pi}_l a_0+O(|s|^3),
\end{multline*}
and, on applying lemma \ref{coherence}, we have that equation \eref{eq1b} is equivalent to
\begin{equation}
\label{amp1}
(I-\widetilde{\pi})a_0 = O(|s|^3).
\end{equation}
Analogously, for the inhomogeneous equation \eref{eq2b}, we find that the component $\widetilde{\pi} a_1$ is arbitrary since $\sigma_{L_0} \widetilde{\pi} a_1 =O(|s|^3)$, so we can set $\widetilde{\pi} a_1 =0$ and $a_1 = (I-\widetilde{\pi})a_1$. Then, by applying the extended projectors $\widetilde{\pi}$ and $I-\widetilde{\pi}$ to the inhomogeneous equation we get the necessary condition
\begin{equation}
\label{amp2}
\widetilde{\pi} \big(L_0(t,x,\partial)a_0 + B(t,x)a_0\big) =O(|s|),
\end{equation}
since $\widetilde{\pi}\sigma_{L_0}a_1 =O(|s|^3)$, and the algebraic equation
\begin{equation}
\label{amp3}
(I-\widetilde{\pi})\sigma_{L_0}(t,x,d\phi) (I-\widetilde{\pi})a_1 + (I-\widetilde{\pi}) \big(L_0(t,x,\partial)a_0 + B(t,x)a_0\big) =O(|s|).
\end{equation}
Equations \eref{amp1}-\eref{amp3} should be satisfied in the neighbourhood $\mathcal{O}$ where coordinates $(t,r,s)$ are defined.

On considering first equation \eref{amp1}, we write
\begin{equation*}
\widetilde{\pi}(t,x,d_x\phi) = \pi(t,r) + \sum \pi_i(t,r) s_i + \tfrac{1}{2}\sum \pi_{ij}(t,r) s_i s_j + O(|s|^3),
\end{equation*}
where $\pi(t,r)=\pi(t,x,d_x \varphi(t,x))|_{R}$, whereas $\pi_i(t,r) = \partial_{s_i}\widetilde{\pi}|_R$ and $\pi_{ij} = \partial_{s_j s_j}\widetilde{\pi}|_R$. The Taylor expansion for the amplitude show that equation \eref{amp1} determines the first three coefficients of the expansion only. Clearly, the space of second-degree polynomials in $s$ with coefficients in $C^\infty(R)$ constitutes a $C^\infty(R)$-module. We now show that the solutions of \eref{amp1} are one-to-one to functions in $C^\infty(R;\C^N)$ satisfying the appropriate polarization condition. It is worth noting that $\pi(t,r)$ is the projector $\pi(t,x,\xi)$ restricted to the isotropic manifold $\Lambda$, locally parametrized by $(t,r)$.

\begin{proposition}
\label{furbata}
A function $a_0\in C^\infty(\ol{\Omega};\C^N)$ solves \eref{amp1} if and only if its restriction to a neighbourhood $\mathcal{O}$ of $R$ amounts to $M(t,r,s)a(t,r)$ where $a \in C^\infty(R;\C^N)$ satisfies the polarization condition $(I-\pi(t,r))a(t,r)=0$ and
\begin{equation*}
M=I+\sum M_i(t,r) s_i + \frac{1}{2}\sum M_{ij}(t,r)s_i s_j, 
\end{equation*}
$M_i, M_{ij} \in C^\infty(R;\mathrm{End}(\C^N))$, $\pi(t,r) M_i(t,r) \pi(t,r) = \pi(t,r) M_{ij}(t,r) \pi(t,r) =0$.
\end{proposition}

First, it is useful to prove the following lemma.

\begin{lemma}
\label{simple1}
With the notations given above we have $\pi(t,r) \pi_i(t,r) \pi(t,r) =0$ and
\begin{equation*}
\pi(t,r) (\pi_i(t,r) \pi_j(t,r) + \pi_j(t,r) \pi_i(t,r) + \pi_{ij}(t,r)) \pi(t,r) =0.
\end{equation*}
\end{lemma}

\begin{proof}
Since $\widetilde{\pi}^2 - \widetilde{\pi} =O(|s|^3)$, i.e.,  $\partial_{s_i}(\widetilde{\pi}^2-\widetilde{\pi})|_R =0$ and $\partial_{s_i}\partial_{s_j}(\widetilde{\pi}^2-\widetilde{\pi})|_R =0$, the claim follows on performing explicitly the derivatives and multiplying by $\pi(t,r)$ both on the right and on the left.
\end{proof}

\begin{proof}[Proof of proposition \ref{furbata}]
Let $a, a_i, a_{ij}$ be the coefficients of the expansion of $a_0$ near $R$ modulo $O(|s|^3)$. Then, we have
\begin{multline*}
(I-\widetilde{\pi}) a_0 = (I-\pi(t,r)) a + \sum \big[(I- \pi(t,r))a_i - \pi_i(t,r) a  \big] s_i \\
+\frac{1}{2} \sum \big[ (I- \pi(t,r))a_{ij} - \pi_i(t,r) a_j - \pi_j(t,r) a_i - \pi_{ij}(t,r) a  \big] s_i s_j + O(|s|^3).
\end{multline*}
We see that $a_0$ solves equation \eref{amp1} if and only if
\begin{gather*}
(I-\pi(t,r))a=0,\quad (I-\pi(t,r))a_i = \pi_i(t,r) a, \\
(I-\pi(t,r))a_{ij} = \pi_i(t,r) a_j + \pi_j(t,r) a_i + \pi_{ij}(t,r)a.
\end{gather*} 
The first equation is identically satisfied in virtue of the hypotheses. Lemma \ref{simple1} provides the solvability conditions for the remaining equations the solution of which is thus readily found with the result that $M_i(t,r) = \pi_i(t,r)$ and $M_{ij}(t,r)=\pi_i(t,r) \pi_j(t,r) + \pi_j(t,r) \pi_i(t,r) + \pi_{ij}(t,r)$. Finally, again by means of lemma \ref{simple1}, one can see that $\pi(t,r) M_i(t,r) \pi(t,r) = \pi(t,r) M_{ij}(t,r) \pi(t,r)=0$.
\end{proof}

There is a natural smooth extension of $Ma$ to the whole domain $\ol{\Omega}$ given by
\begin{equation*}
a_0(t,x) = \pi \ul{a} + i \sum (\partial_{\xi_i}\pi)\pi \ul{a} \chi_i -\frac{1}{2}\sum (\partial_{\xi_i}\pi \partial_{\xi_j}\pi + \partial_{\xi_j} \pi \partial_{\xi_i} \pi + \partial_{\xi_i\xi_j} \pi) \pi \ul{a} \chi_i \chi_j,
\end{equation*}
where $\ul{a}(t,x)$ is any smooth extension of $a(t,r)$ to a compact neighbourhood of $R$, $\chi_i(t,x) = \partial \chi(t,x)/\partial x_i$ and the derivatives of $\pi$ are to be evaluated at $(t,x,d_x\varphi)$. This follows by direct substitution into \eref{amp1} and by using the following identities that can be proved as lemma \ref{simple1}.

\begin{lemma}
\label{simple}
If $\pi(x)$ is a projector-valued $C^2$-function of $x \in \R^k$, then $\pi \partial_i \pi \pi =0$, and $\pi (\partial_i \pi \partial_j \pi + \partial_j \pi \partial_i \pi + \partial_{ij}\pi) \pi =0$, where $\partial_i = \partial /\partial x_i$ and $\partial_{ij}= \partial^2 /\partial x_i \partial x_j$.
\end{lemma}

Let us now consider \eref{amp2} with $a_0$ given above. First we note that,
\begin{equation*}
\partial_t a_0 = \partial_t (\pi\ul{a}) + O(|s|), \quad \partial_{x_i}a_0 = \partial_{x_i}(\pi \ul{a}) + i \sum_j \partial_{\xi_j}\pi \pi \ul{a} \ \frac{\partial^2 \chi}{\partial x_i \partial x_j} + O(|s|),
\end{equation*}
hence,
\begin{equation*}
L_0(t,x,\partial) a_0 = L_0(t,x,\partial)\pi\ul{a} + i\Big(\sum_{i,j}  \frac{\partial^2 \chi}{\partial x_i \partial x_j} A_i \partial_{\xi_j}\pi \pi\Big)\ul{a} + O(|s|),
\end{equation*}
and equation \eref{amp2} reads
\begin{equation*}
\pi L_0(t,x,\partial)\pi\ul{a} + \frac{i}{2} \Big(\sum_{i,j}  \frac{\partial^2 \chi}{\partial x_i\partial x_j}  \pi \big(A_i \partial_{\xi_j}\pi + A_j \partial_{\xi_i}\pi\big) \pi \Big)\ul{a} + \pi B\pi \ul{a} = O(|s|).
\end{equation*}
The differential term in this equation can be further simplified on writing 
\begin{equation*}
\pi L_0\pi\ul{a} = \pi \mathcal{L}_V \ul{a} + \pi (L_0\pi)\ul{a},
\end{equation*}
where where $\mathcal{L}_V$ is the Lie derivative along the vector field $V=\partial_t +d_\xi\lambda(t,x,d_x\varphi)$ and we have used the identity 
\begin{equation*}
\pi A_k \pi = \pi \frac{\partial A}{\partial \xi_k} \pi = \pi \Big(\frac{\partial (A\pi)}{\partial \xi_k} - A\frac{\partial \pi }{\partial \xi_k} \Big) = \pi\Big(\frac{\partial \lambda}{\partial \xi_k} \pi + (\lambda -A)\frac{\partial \pi}{\partial \xi_k}\Big) = \frac{\partial \lambda}{\partial \xi_k} \pi.
\end{equation*}

As for the second term, the symmetric part of $A_i \partial_{\xi_j}\pi$ can be computed as follow. From one hand, we have
\begin{equation*}
\frac{\partial^2 (A\pi)}{\partial \xi_i \partial \xi_j} = \Big(\frac{\partial A}{\partial \xi_i}\frac{\partial \pi}{\partial \xi_j} + \frac{\partial A}{\partial \xi_j}\frac{\partial \pi}{\partial \xi_i} \Big) + A \frac{\partial^2 \pi}{\partial \xi_i \partial \xi_j},
\end{equation*}
where $A = \sum A_i \xi_i$, thus, $A_i = \partial A/\partial \xi_i$ and $\partial^2 A/\partial \xi_i \partial \xi_j =0$. On the other hand,
\begin{equation*}
\frac{\partial^2 (A\pi)}{\partial \xi_i \partial \xi_j} = \frac{\partial^2 (\lambda\pi)}{\partial \xi_i \partial \xi_j} = \frac{\partial^2 \lambda}{\partial \xi_i \partial \xi_j} \pi +\Big(\frac{\partial \lambda}{\partial \xi_i}\frac{\partial \pi}{\partial \xi_j} + \frac{\partial \lambda}{\partial \xi_j}\frac{\partial \pi}{\partial \xi_i} \Big) + \lambda \frac{\partial^2 \pi}{\partial \xi_i \partial \xi_j},
\end{equation*}
so that
\begin{equation*}
\big(A_i\partial_{\xi_j}\pi + A_j \partial_{\xi_i}\pi\big) = \frac{\partial^2 \lambda}{\partial \xi_i \partial \xi_j} \pi +\Big(\frac{\partial \lambda}{\partial \xi_i}\frac{\partial \pi}{\partial \xi_j} + \frac{\partial \lambda}{\partial \xi_j}\frac{\partial \pi}{\partial \xi_i} \Big) + (\lambda -A) \frac{\partial^2 \pi}{\partial \xi_i \partial \xi_j}.
\end{equation*}
Finally, in virtue of lemma \ref{simple}, we find
\begin{equation*}
\pi \big(A_i \partial_{\xi_j}\pi + A_j \partial_{\xi_i}\pi\big) \pi = \frac{\partial^2\lambda}{\partial \xi_i \partial \xi_j} \pi.
\end{equation*}
Since $V$ is tangent to $R$ we find that equation \eref{amp2} is satisfied if and only if $a(t,r) \in C^\infty(R;\C^N)$ solves the transport equation
\begin{equation}
\label{transport-equation}
\left\{ \begin{aligned}
& (I - \pi)a = 0,\\
& \pi \mathcal{L}_V  a + \pi(L_0\pi + B) a + ig_\chi a = 0,
\end{aligned}\right.
\end{equation}
where all the coefficient should be evaluated on $R$ and
\begin{equation*}
g_\chi(t,r) = \frac{1}{2} \sum_{i,j} \frac{\partial^2\chi}{\partial x_i \partial x_j} \frac{\partial^2\lambda}{\partial \xi_i \partial \xi_j}\bigg|_{R},
\end{equation*}
introduces a phase-shift in the amplitude due to the wave field localization. Apart from such a phase-shift effect this is exactly the geometric optics transport equation evaluated on $R$. We also note that, since $\partial a_0|_R$ is independent on the choice of the smooth extension $a_0$, the transport equation is also independent on such choice.

The transport equation \eref{transport-equation} allows us to obtain a solution for $a$ by integrating along the integral lines of the vector field $V$, that are parametrized by $R^o = R \cap \{t=0\}$, then, we get a solution for $a_0$ given in terms of $a$ by proposition \ref{furbata}.

Finally, we have to solve \eref{amp3} which is purely algebraic and one readily see that, with $a_1 = (I-\widetilde{\pi})a_1$, it is equivalent to
\begin{equation*}
\big[(I-\widetilde{\pi})\sigma_{L_0} a_1 + (I-\widetilde{\pi})(L_0 a_0 + Ba_0)\big]_{|R} =0.
\end{equation*}
Since
\begin{equation*}
(I-\widetilde{\pi})\sigma_{L_0} = \sum_{l; \lambda_l \not =\lambda} i\big(\partial_t \phi + \widetilde{\lambda}_l(t,x,d_x\phi)\big) \widetilde{\pi}_l(t,x,d_x\phi),
\end{equation*}
lemma \ref{coherence} ensure that such an operator is invertible on a neighbourhood of $R$ and denoting by $Q(t,r)$ its inverse evaluated on $R$ we can set $a_1(t,x)$ equal to any smooth extension of
\begin{equation}
\label{corrector}
-Q(t,r) \big[(I-\widetilde{\pi})(L_0 a_0 + Ba_0)\big]_{|R},
\end{equation}
to a compact neighbourhood of $R$.

\section{Construction of the Complex Geometric Optics Solution}
\label{proof1}

Now we can construct the complex geometric optics solution of the Cauchy problem for \eref{L} provided that assumptions \ref{hyperbolicity}-\ref{ref-man} hold true together with condition \ref{pol}. 

First, we have to solve the Hamilton's equations and obtain the isotropic manifolds $\Lambda_1,\ldots\Lambda_m$ along with the corresponding projections $R_1,\ldots,R_m$ that give the reference manifold $R = \bigcup_\mu R_\mu$; the existence of such geometric objects is ensured by assumption \ref{ref-man} in section \ref{results}. 

Then, we make use of the construction of section \ref{gcee} in order to obtain the coordinates patches $\mathcal{O}_{\mu,\ell}$ on the basis of a finite covering $\{\mathcal{R}_{\mu,\ell}\}_\ell$ of $R^o_\mu$; that exists since, in view of assumption \ref{Rzero}, $R^o_\mu$ is a closed subset of the compact set $\ol{X^o}$, hence, it is compact. In each neighbourhood $\mathcal{O}_{\mu,\ell}$ we have coordinates $(t,r,s) \in [0,T]\times \mathcal{O}_r \times \mathcal{O}_s$. 

Next, we obtain the approximate solutions $\phi_\mu$ to the Cauchy problems,
\begin{equation*}
\partial_t \phi_\mu + \widetilde{\lambda}_{l(\mu)}(t,x,d_x\phi_\mu) =0,\qquad \phi_{\mu|t=0}(x)=\psi_\mu(x),
\end{equation*}
with $l(\mu)$ being given in condition \ref{pol}. In the statement of proposition \ref{main-linear} we have assumed $d^2_s\chi_\mu|_{R^o}(x) >0$ and, according to remark \ref{riccati-rem}, it is $\chi_\mu(t,r,s) \geq c_\mu |s|^2$ for some constants $c_\mu>0$: this allows us to apply lemma \ref{maslov-estimate2-small} with $q=2$. The phases $\phi_\mu$ are defined in $\mathcal{O}_{\mu} = \bigcup_\ell \mathcal{O}_{\mu,\ell}$ up to a remainder of $O(|s|^3)$, therefore, they are better understood as a representative of an equivalence class for the following equivalence relation in $C^\infty(\ol{\Omega};\C^m)$,
\begin{equation*}
\begin{aligned}
&\text{every two $\phi,\phi' \in C^\infty(\ol{\Omega};\C^m)$ are equivalent if and only if each component}\\ 
&\text{$\phi_\mu$ and $\phi_\mu'$ have the same second-degree Taylor polynomial in the variable $s$}\\
&\text{near the submanifold $R_\mu$}.
\end{aligned}
\end{equation*}

We pick a representative $\phi(t,x)$ such that $\mathrm{Im} \phi_\mu >0$ outside $\bigcup_{\ell} \mathcal{O}_{\mu,\ell}$. Given such a representative, we can apply the results of section \ref{amplitudes-linear} in order to obtain a solution $a^{(0)}_\mu$ and $a^{(1)}_\mu$ of \eref{eq1} and \eref{eq2} with $q=2$. Such solutions are defined in the neighbourhood $\mathcal{O}_\mu$ modulo a remainders $O(|s|^3)$ and $O(|s|)$, respectively; in analogy to the complex phases, this defines an equivalence relation in $C^\infty(\ol{\Omega};\C^N)$ of functions with the same Taylor polynomial near $R_\mu$.

Finally, we can construct the (lowest order) complex geometric optics solution \eref{cgo} which is therefore defined modulo the foregoing equivalence relations.

\begin{proof}[Proof of proposition \ref{main-linear}]
The final part of the argument relies on the Maslov's estimates proved in lemmas \ref{maslov-estimate1} and \ref{maslov-estimate2-small} as well as on a partition of unity subordinated to the finite covering of $R^o$ with open sets $\mathcal{R}_{\mu,\ell}^o \subseteq R^o_\mu$. 

For each open set $\mathcal{R}_{\mu,\ell}^o$ let us consider the neighbourhood $\mathcal{O}_{\mu,\ell} \subseteq \ol{\Omega}$ where coordinates $(t,r,s)$ are defined. Assertion (iv) of assumption \ref{ref-man} allows us take $\mathcal{O}_{\mu,\ell}$ so small that $\mathcal{O}_{\mu,\ell} \cap \mathcal{O}_{\nu,\ell'} = \emptyset$ if $\mu \not = \nu$. The open sets $\mathcal{O}^o_{\mu,\ell} = \mathcal{O}_{\mu,\ell}\cap X^o$ satisfies the hypotheses of the partition of unity at $R^o$ \cite[Theorem 1.4.5]{H1}. Therefore, one can find functions $\omega^o_{\mu,\ell} \in C^\infty_0(\mathcal{O}^o_{\mu,\ell})$ such that $\omega^o_{\mu,\ell}\geq 0$ and $\sum_{\mu,\ell} \omega^o_{\mu,\ell} \leq 1$ with equality in a neighbourhood of $R^o$.

By using coordinates $(t,r,s)$, we define the functions $\omega_\mu(t,r,s) = \omega^o_\mu (r,s)$ and we see that $\omega_{\mu,\ell} \in C^\infty(\mathcal{O}_{\mu,\ell})$ and, for every $t \in [0,T]$, 
\begin{equation*}
\begin{aligned}
&\qquad\qquad\qquad\text{$\omega_{\mu,\ell}(t,\cdot) \in C^\infty_0(\mathcal{O}_{\mu,\ell} \cap X^t)$ and}\\
&\text{ $ \sum_{\ell} \omega_{\mu,\ell}(t,\cdot) = 1$ in a neighbourhood of $R^t_{\mu} = R_{\mu} \cap X^t$}.
\end{aligned}
\end{equation*}
The sum is over $\ell$ only as, in a neighbourhood of $R_\mu^t$, we have $\omega_{\nu,\ell}=0$ if $\nu \not = \mu$.

Proof of a). Let us write $1 = \big(1- \sum_{\ell} \omega^o_{\mu,\ell}\big) + \sum_{\ell}\omega^o_{\mu,\ell}$ so that $(1-\sum_{\ell} \omega^o_{\mu,\ell})$ is supported away from $R^o_\mu$ whereas $\omega^o_{\mu,\ell}$ is supported in $\mathcal{O}^o_{\mu,\ell}$. More specifically we have,
\begin{multline*}
\big|h_\mu e^{i\psi_\mu/\ep} - v^{\ep}_{\mu |t=0}\big|  \leq \sum_{\ell} \big|\omega_{\mu,\ell}^o \big(h_\mu e^{i\psi_\mu/\ep} - v^{\ep}_{\mu |t=0}\big)\big| \\
 + \big|\big(1-\sum_{\ell} \omega^o_{\mu,\ell} \big)\big(h_\mu e^{i\psi_\mu/\ep} - v^{\ep}_{\mu |t=0}\big) \big|,
\end{multline*}
and, in virtue of lemma \ref{maslov-estimate1} the last term is $\leq C_{\mu,k} \ep^k$ for every $k \in \N$, with the $\sup$ being computed over $\ol{S} = \ol{\mathrm{supp}(1-\sum_\ell \omega^o_{\mu,\ell})} \subset \ol{X^o}$. As for the first term we exploit the Taylor's formula for $\psi_\mu$ and, on recalling that $\phi_{\mu|t=0}$ amounts just to the second degree Taylor polynomial of $\psi_\mu$, we get
\begin{equation*}
e^{i\psi_\mu(r,s)/\ep} = e^{i \big[\phi_{\mu|t=0}(r,s) + \sum_{|\alpha|=3} s^\alpha \psi_{\mu,\alpha}(r,s)\big]/\ep} = e^{i\phi_{\mu|t=0}(r,s)/\ep} \big[1 + \varrho^\ep_\mu(r,s,\ep) \big],
\end{equation*}
where the remainder $\varrho^\ep_\mu(r,s)$ has been obtained from the fundamental theorem of calculus, 
\begin{equation*}
\varrho^\ep_\mu(r,s,\ep) = \frac{i}{\ep} \sum_{|\alpha|=3} s^\alpha \psi_{\mu,\alpha}(r,s) \int_{0}^1 e^{it \sum_{|\alpha|=3} s^\alpha \psi_{\mu,\alpha}(r,s) /\ep} dt,
\end{equation*}
and the integral is uniformly bounded by $1$. Therefore, we have
\begin{multline*}
\big| \omega_{\mu,\ell}^o \big( h_\mu e^{i\psi_\mu/\ep} - v^{\ep}_{\mu |t=0}\big)\big| \leq \big| \omega^o_{\mu,\ell} \big(h_\mu - a^{(0)}_{\mu|t=0} \big) e^{i\phi_{\mu|t=0}/\ep} \big| \\
+ \ep \big|\omega^o_{\mu,\ell} a^{(1)}_{\mu|t=0} e^{i\phi_{\mu|t=0}/\ep}\big| + \big|\omega^o_{\mu,\ell} \varrho^\ep_\mu e^{i\phi_{\mu|t=0}/\ep}\big|.
\end{multline*}
The first term is of $O(\ep^{\frac{1}{2}})$ in virtue of lemma \ref{maslov-estimate2-small} applied at $t=0$ and with the compact set $K = \mathrm{supp}(\omega^o_{\mu,\ell})$, as $h_\mu$ and $a_{\mu|t=0}^{(0)}$ are the same when evaluated on $R^o_\mu$. The second term is $O(\ep)$ uniformly on $\ol{X^o}$, whereas the third term can be estimated by
\begin{equation*}
\big|\omega^o_{\mu,\ell} \varrho^\ep_\mu e^{i\phi_{\mu|t=0}/\ep}\big| \leq  \frac{1}{\ep} \sum_{|\alpha|=3} \big|s^\alpha e^{i\phi_{\mu|t=0}/\ep}\big| \sup |\omega^o_{\mu,\ell} \psi_{\mu,\alpha}|\leq \ep^{\frac{1}{2}}\ \text{const.},  
\end{equation*}
again, uniformly in $\ol{X^o}$, where we have used the fact that
\begin{equation*}
|s^{\alpha} e^{i\phi_{\mu|t=0}/\ep}\big| \leq \ep^{\frac{3}{2}} |v^{\alpha} e^{-c v^2}|, \quad v=s/\ep^{\frac{1}{2}}, \quad |\alpha|=3,
\end{equation*}
and this proves assertion a).

Proof of b). We write $Lv^\ep_\mu = (1-\sum_\ell \omega_{\mu,\ell})Lv^\ep_\mu + \sum_\ell \omega_{\mu,\ell} Lv^\ep_\mu$, where $L = L_0+B$, and we have, cf. equation \eref{eq},
\begin{equation*}
Lv^\ep_\mu = b^\ep_\mu e^{i\phi_\mu/\ep},\quad b_\mu^\ep = \ep^{-1} \sigma_{L_0} a^{\ep}_\mu + La^{\ep}_\mu,
\end{equation*}
with
\begin{equation*}
\big|(1-\sum_{\ell} \omega_{\mu,\ell}) b_\mu^\ep e^{i\phi_\mu/\ep} \big| \leq C_{k,\mu} \ep^k,
\end{equation*}
uniformly in $\ol{\Omega}$ and for every $k \in \N$ in view of lemma \ref{maslov-estimate1}. Moreover,
\begin{multline*}
\big|\omega_{\mu,\ell} \big[ \ep^{-1} \sigma_{L_0} a_\mu^\ep + L a_\mu^\ep \big]e^{i\phi_\mu/\ep}\big| \leq \ep^{-1} |\omega_{\mu,\ell} \sigma_{L_0} a_\mu^{(0)}e^{i\phi_\mu/\ep}\big| \\
+ \big| \omega_{\mu,\ell} \big( \sigma_{L_0} a_\mu^{(1)} + L_0 a_\mu^{(0)} + Ba_\mu^{(0)}\big)e^{i\phi_\mu/\ep}\big| + \ep \big|\omega_{\mu,\ell} \big( L_0 a_\mu^{(1)} + Ba_\mu^{(1)}\big)e^{i\phi_\mu/\ep}\big|.
\end{multline*}
We know that equations \eref{eq1} and \eref{eq2} hold so that the second degree Taylor polynomial in $s$ of $\omega_\mu \sigma_{L_0} a_\mu^{(0)}$ and the zero-degree Taylor polynomial in $s$ of $\omega_\mu \big( \sigma_{L_0} a_\mu^{(1)} + L_0 a_\mu^{(0)} + Ba_\mu^{(0)}\big)$ vanishes. Hence, we can apply lemma \ref{maslov-estimate2-small} on the compact set $\mathrm{supp}(\omega_{\mu,\ell}) \subset \ol{\Omega}$ so that the first two terms are $O(\ep^{\frac{1}{2}})$ uniformly on the whole domain $\ol{\Omega}$. Finally, the last term is $O(\ep)$.
\end{proof}

\begin{proof}[Proof of corollary \ref{conv}]
For every fixed $\ep \in \R_+$, we know that there exists a unique solution $u^\ep \in C^\infty(\ol{\Omega})$, which satisfies the energy inequality,
\begin{equation*}
\|u^\ep(t)\|_{L^2(X^t)} \leq e^{tK}\|u^\ep(0)\|_{L^2(X^o)},
\end{equation*}
where the constant $K$ depends only on $L^\infty$-norms of $\sum \partial_{x_i}A_i(t,x)$ and $B(t,x)$. In particular, $K$ does not depends on $\ep$ and $\{u^\ep(0);\ep \in (0,\ep_0]\}$ is a bounded family in $L^2(X^o)$, thus, $\{u^\ep;\ep \in (0,\ep_0]\}$, with $u^\ep(t,x)$ extended by zero outside $\ol{\Omega}$, is a bounded family in $C([0,T],L^2(\R^d))$.

Let $v^\ep$ be any representative of the equivalence class in proposition \ref{main-linear} in a common domain of determinacy $\ol{\Omega}$. Then, the difference $w^\ep = u^\ep - v^\ep$ satisfies the problem
\begin{equation}
\label{difference}
\left\{
\begin{aligned}
& L(t,x,\partial) w^\ep(t,x) = f^\ep(t,x) \in C^\infty(\ol{\Omega};\C^N),\\
& w^\ep_{|t=0}(x)=g^\ep(x),
\end{aligned}\right.
\end{equation}
where, in view of proposition \ref{main-linear},
\begin{equation*}
\|g^\ep\|_{L^\infty(X^o)} = \sup |g^\ep(x)| \leq C \ep^{\frac{1}{2}}, \quad \|f^\ep \|_{L^\infty(\Omega)} = \sup |f^\ep(t,x)| \leq C' \ep^{\frac{1}{2}}.
\end{equation*}
In addition, for $0\leq t \leq T$, we have the inhomogeneous version of the energy inequality
\begin{equation}
\label{L2estimate}
\|w^\ep(t)\|_{L^2(X^t)} \leq e^{tK} \|g^\ep\|_{L^2(X^0)} + \int_0^t e^{(t-t')K}\|f^\ep(t')\|_{L^2(X^{t'})} dt', 
\end{equation}
with the same constant $K$ as before. Then, we find
\begin{equation*}
\|w^\ep(t)\|_{L^2(X^t)} \leq a(t)\|g^\ep\|_{L^\infty(\Omega)} + b(t) \|f^\ep\|_{L^\infty(\Omega)} \leq c(t) \ep^{\frac{1}{2}},
\end{equation*}
and $c \in C([0,T];\R_+)$.
\end{proof}

\section*{Acknowledgments}
This work has been supported partly by the CNISM, at the Department of Physics ``A.~Volta'' of the Pavia University (Italy), under the grant ``Propagazione di onde ad alta frequenza in mezzi dispersivi e disomogenei: dalla teoria dei sistemi dinamici alle applicazioni'' and by the Foundation Blanceflor Boncompagni-Ludovisi at the Max-Planck-Institut f\"ur Plasmaphysik (IPP), Garching bei M\"unchen (Germany). I thank the Theory Division of the IPP and, in particular, G.~V.~Pereverzev and E.~Poli for their kind hospitality and collaboration. I am greatly indebted to M.~Bornatici for his continuous encouragement and tutorship. Especially, I wish to thank C.~Dappiaggi for so many discussions, suggestions and for carefully reading the manuscript.


\begin{thebibliography}{0}
\bibitem{R} J.~Rauch and M.~Keel, Lectures on geometric optics, in {\it Hyperbolic equations and frequency interactions} (Park City, UT, 1995), pp.383--466, AMS, Providence, RI.

\bibitem{Ma} V.~P.~Maslov, {\it The Complex-WKB Method for Nonlinear Equations I: Linear Theory} (Birkh\"auser, Boston 1996).

\bibitem{P96} G.~V.~Pereverzev, Paraxial WKB solution of a scalar wave equation, {\it Reviews of Plasma Physics} {\bf 19} (1996) 1--52.

\bibitem{P98} G.~V.~Pereverzev, Beam tracing in inhomogeneous anisotropic plasmas, {\it Phys. of Plasmas} {\bf 5} (1998) 3529--3541.

\bibitem{Kr} Yu.~A.~Kravtsov, {\it Geometrical Optics in Engineering Physics} (Alpha Science, 2005).

\bibitem{KrBe} Yu.~A.~Kravtsov and P.~Berczynski, Gaussian beams in inhomogeneous media: a review, {\it Stud. Geophys. Geod.} {\bf 51} (2007) 1--36.

\bibitem{AR1} D.~Alterman and J.~Rauch, Nonlinear geometric optics for short pulses, {\it J. Differential Equations} {\bf 178} (2002) 437--465.

\bibitem{AR2} D.~Alterman and J.~Rauch, Diffractive nonlinear geometric optics for short pulses, {\it SIAM J. Math. Anal.} {\bf 34} (2003) 1477--1502. 

\bibitem{BB} V.~M.~Babi\v{c} and V.~S.~Buldyrev, {\it Short-Wavelength Diffraction Theory} (Springer-Verlag, Berlin 1991).

\bibitem{Lax} P.~D.~Lax, Asymptotic solutions of oscillatory initial value problems, {\it Duke Math. J.} {\bf 24} (1957) 627--646.

\bibitem{AMR} R.~Abraham, J.~E.~Marsden and T.~Ratiu, {\it Manifolds, Tensor Analysis, and Applications} Second edition (Springer, New York 1988).

\bibitem{Ka} T.~Kato, {\it Perturbation Theory for Linear Operator} (Springer, Berlin 1995).

\bibitem{H1} L.~H\"ormander, {\it The Analysis of Partial Differential Operators I: Distribution Theory and Fourier Analysis} (Springer-Verlag, Berlin 2003).

\bibitem{H3} L.~H\"ormander, {\it The Analysis of Partial Differential Operators III: Pseudo-Differential Operators} (Springer-Verlag, Berlin 1985).

\end{thebibliography}
\end{document}